\documentclass[iop]{emulateapj}

\usepackage{natbib}
\usepackage{amsmath}
\usepackage{hyperref}

\newcommand{\eps}{\varepsilon}

\begin{document}

\title{Numerical sunspot models: Robustness of photospheric velocity and 
  magnetic field structure}
\shorttitle{Numerical sunspot models}

\author{M. Rempel}

\shortauthors{M. Rempel}
 
\affil{High Altitude Observatory,
    NCAR, P.O. Box 3000, Boulder, Colorado 80307, USA}

\email{rempel@ucar.edu}

\begin{abstract}
MHD simulations of sunspots have successfully reproduced many
aspects of sunspot fine structure as consequence of magneto convection in
inclined magnetic field. We study how global sunspot properties 
and penumbral fine structure depend on the magnetic top boundary condition
as well as on grid spacing. The overall radial extent of the penumbra is 
subject to the magnetic top boundary condition. All other aspects of
sunspot structure and penumbral fine structure are resolved at an acceptable 
level starting from a grid resolution of $48\,[24]$~km (horizontal [vertical]). 
We find that the amount of inverse polarity flux and the overall 
amount of overturning convective motions in the penumbra are robust 
with regard to both, resolution and boundary conditions. At photospheric
levels Evershed flow channels are strongly magnetized. We discuss in detail
the relation between velocity and magnetic field structure in the photosphere
and point out observational consequences.  
\end{abstract}


\received{}
\accepted{}

\maketitle
\section{Introduction}
Numerical simulations of sunspot fine structure have seen substantial 
improvements over that past 5 years. \citet{Schuessler:Voegler:2006}
presented the first realistic MHD simulations of umbral dots, followed
by a series of simulations in 'slab-geometry':
\citet{Heinemann:etal:2007,Scharmer:etal:2008,Rempel:etal:2009} addressing
the transition from umbra to penumbra and gave the first 
insight into the origin of penumbra fine structure and the Evershed flow. 
\citet{Kitiashvili:etal:2009} focused on a magneto convection simulation in 
strongly inclined field representative of penumbral conditions and studied
the influence of field strength and inclination on the presence horizontal 
outflows. \citet{Rempel:etal:Science,
Rempel:2011:IAU,Rempel:2011,Rempel:2011:moat} focused on full circular 
sunspots with increasing realism, with the best resolved simulations such 
as those shown in \citet{Rempel:2011:IAU} reaching and exceeding the 
richness of detail seen in the best available observations today. As described
in detail in \citet{Rempel:2011:IAU,Rempel:2011} these simulations
reproduce the key aspects of penumbral fine structure as
inferred from high resolution observations
\citep[see, e.g.][]{Scharmer:etal:2002,Langhans:etal:2005,
Rimmele:Marino:2006,Ichimoto:etal:2007,Ichimoto:etal:2007:sc,
Langhans:etal:2007,Scharmer:etal:2007,Rimmele:2008,Franz:Schlichenmaier:2009,
BellotRubio:etal:2010,Franz:2011} and recent reviews by 
\citet{Solanki:2003,Thomas:Weiss:2004,Thomas:Weiss:2008,
Rempel:Schlichenmaier:2011,Borrero:Ichimoto:2011}.

Simulations are still far from perfect. The numerical grid resolution might be
marginal for resolving relevant details of the penumbra, the initial condition 
in terms of a monolithic axisymmetric field is very artificial, and there is 
a strong influence from the (unavoidable) imposed boundary conditions.
Therefore, in this paper we investigate numerical resolution and boundary 
conditions. We also deepen our discussion on the flow field and magnetic 
structure in the observable layers of the penumbra. 
At the bottom boundary most of the sunspot simulations impose a fixed 
magnetic field to prevent a rapid decay of the sunspot. 
\citet{Rempel:2011:moat} showed that this constraint can be relaxed in 
sufficiently deep domains where convective time scales reach several days, 
but not in shallow $\sim 6$~Mm deep domains typically used in simulations 
that address sunspot fine structure. Here we investigate how
the magnetic top boundary condition as well as grid resolution influence
details of penumbral fine structure, while we keep the initial state as
well as the bottom boundary unchanged. We deepen the 
analysis presented in \citet{Rempel:2011} to test the numerical robustness of 
their findings. In addition we expand the discussion to address three aspects 
that have been subject to major controversy in recent years:

(1) {\em The convective nature of the penumbra:} 
Vigorous convection (about half the strength of granulation) is the key
process in MHD simulations leading to penumbral fine structure. Indirect
observational evidence for overturning motions is derived from 'twisting' 
motions \citep{Ichimoto:etal:2007:sc,Bharti:etal:2010} or
correlation tracking that shows the convergence of tracers toward the
edge of filaments similar to granulation \citep{Marquez:etal:2006}. 
\citet{Zakharov:etal:2008} found evidence for convective roles in filaments
oriented parallel to the solar limb and estimated that the observed velocities
are sufficient to explain the penumbral brightness.
Direct evidence for convective motions has been questioned by some authors
\citep{Franz:Schlichenmaier:2009,BellotRubio:etal:2010}, while it was found
by others in high resolution observations
\citep{Sanchez-Almeida:etal:2007,Joshi:etal:2011,Scharmer:etal:2011:sci,
Scharmer:Henriques:2011}. \citet{Joshi:etal:2011} found convective 
downflows mostly in the inner penumbra using the rather deep forming CI 5380
line. \citet{Scharmer:etal:2011:sci,Scharmer:Henriques:2011} found consistent
results using the CI 5380 and FeI 6301 lines in terms of an anisotropic
convection pattern with a correlation between intensity and vertical flow
velocity not very different from quiet sun. Such correlations were also
previously indicated in the analysis of \citet{Sanchez-Almeida:etal:2007}.

(2) {\em The presence of opposite polarity magnetic flux throughout the 
penumbra and its relation to downward directed mass flux:}
While inverse polarity flux is certainly present in the outer penumbra
\citep{DelToro:etal:2001}, it is debated if this is also the case in the 
inner penumbra.
\citet{Sanchez-Almeida:2005:sunspot} found substantial amount of opposite
polarity flux throughout the penumbra, accounting to about $58\%$ of the
unsigned magnetic flux of the penumbra, \citet{Langhans:etal:2005} did
not find strong evidence in magnetogram data, possible due to resolution
effects. Recently \citet{Franz:2011} found with {\em Hinode} data that about
$40\%$ of downflows in the outer penumbra are in regions with opposite polarity
flux and consider this a lower limit.

(3) {\em The magnetization of the Evershed flow:} 
Spectropolarimetric observations point to a flow in regions
with a substantial magnetic field \citep[see, e.g.][]{Ichimoto:etal:2008,
Borrero:2009,Borrero:Solanki:2008,Borrero:Solanki:2010},
see also discussions in \citet{Brummell:etal:2008,Thomas:2010}.
Some models suggest that penumbral filaments are
mostly field free 'gaps' \citep{Spruit:Scharmer:2006,Scharmer:Spruit:2006}.
A similar conclusion was also derived from an interpretation of
'twisting filament motions' as fluting instability \citep{Spruit:etal:2010},
leading to an upper bound for magnetic field in flow channels of about
$300$~G. MHD simulations however point toward strong $1-2$ kG
field in flow channels, in the outer penumbra the field is even enhanced
in flow channels compared to the background \citep{Rempel:2011}.

\section{Numerical setup}
\label{sect:setup}
The numerical models were computed with an expanded version of the {\em MURaM}
radiative MHD code \citep{Voegler:etal:2005}. We introduced an artificial 
limitation of Alfv{\'e}n velocities to $60$~km s$^{-1}$ in order to prevent 
severe CFL time step constraints and implemented a new numerical diffusivity 
approach, both are described in \citet{Rempel:etal:2009}. For the
simulations presented here with grid sizes of up to $4.8$ billion grid points
we did additional performance enhancements including a major rewrite of IO 
and improved the scalability of the code to up to $24576$ cores. Most of the 
simulations presented here were computed in the $1024-12288$ core range.

We present here a series of numerical sunspot models which were all computed in 
a domain with a size of $49.152\times 49.152\times 6.144$ Mm$^3$. All models 
were initialized with an axisymmetric self-similar magnetic field configuration
described in Appendix A. For the models presented here we used the parameters
$B_0=6.4$ kG, $R_0=8.2$ Mm, and $z_0=6.4185$ Mm, leading to a sunspot with an 
initial flux of  $\Phi_0=1.2\cdot 10^{22}$ Mx, a field strength of $B_0=6.4$ kG
at the bottom of our domain ($z=0$ Mm), dropping to $2.56$ kG at the top 
($z=6.144$ Mm).

While keeping this initial condition fixed we explore the dependence of the 
resulting sunspots on the top boundary condition and grid resolution.

We use in the horizontal direction periodic boundaries, the bottom boundary
condition is open for convective flows in regions with $\vert B\vert<2.5$ kG, 
but closed
otherwise. The open boundary condition imposes a symmetric mass flux (all three 
components) in the ghost cells and extrapolates the gas pressure such that its
value at the boundary is fixed. In outflow
regions the entropy is determined by the upstream values from within the 
computational domain, in inflow regions the entropy is set to a fixed value 
that leads to the correct solar energy flux under quiet Sun conditions. The 
closed boundary condition in regions with $\vert B\vert>2.5$ kG is implemented 
through 
an antisymmetric mass flux. The top boundary condition is closed (vertical
mass flux antisymmetric) and stress free for horizontal motions. 

\begin{figure*}
  \centering 
  \resizebox{0.95\hsize}{!}{\includegraphics{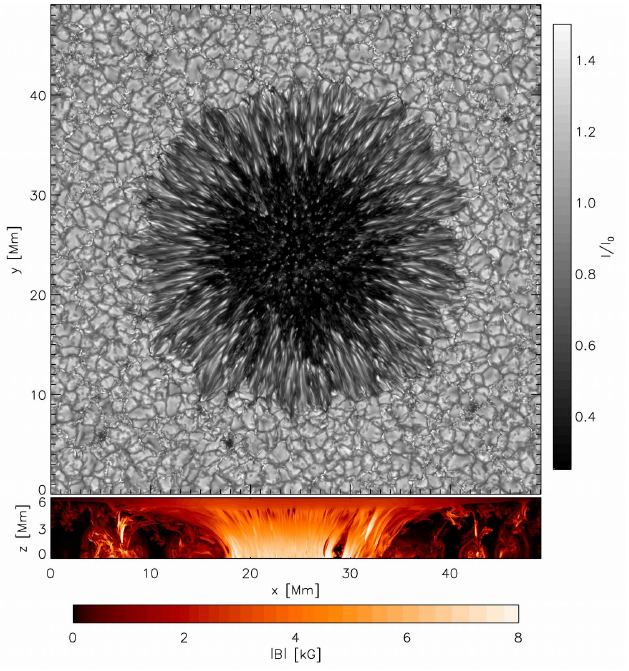}}
  \caption{Top: bolometric intensity, bottom: magnetic field strength
    on a vertical cut through the center of the sunspot. Displayed is
    a snapshot from our highest resolution simulation ($16\,[12]$ km).
    An animation is provided with the online version.
  }
  \label{fig:f0}
\end{figure*}

\begin{figure*}
  \centering 
  \resizebox{0.95\hsize}{!}{\includegraphics{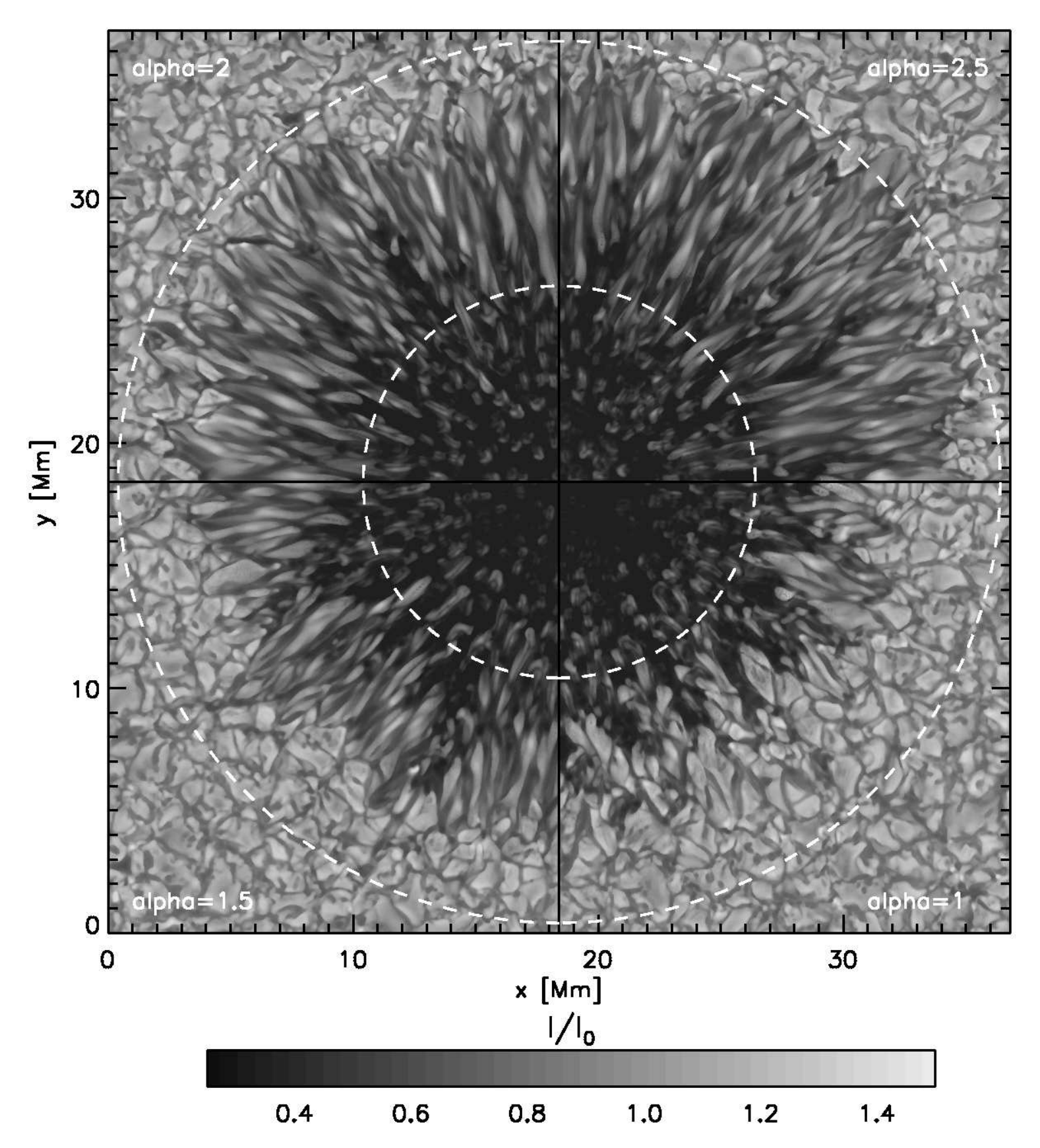}
    \includegraphics{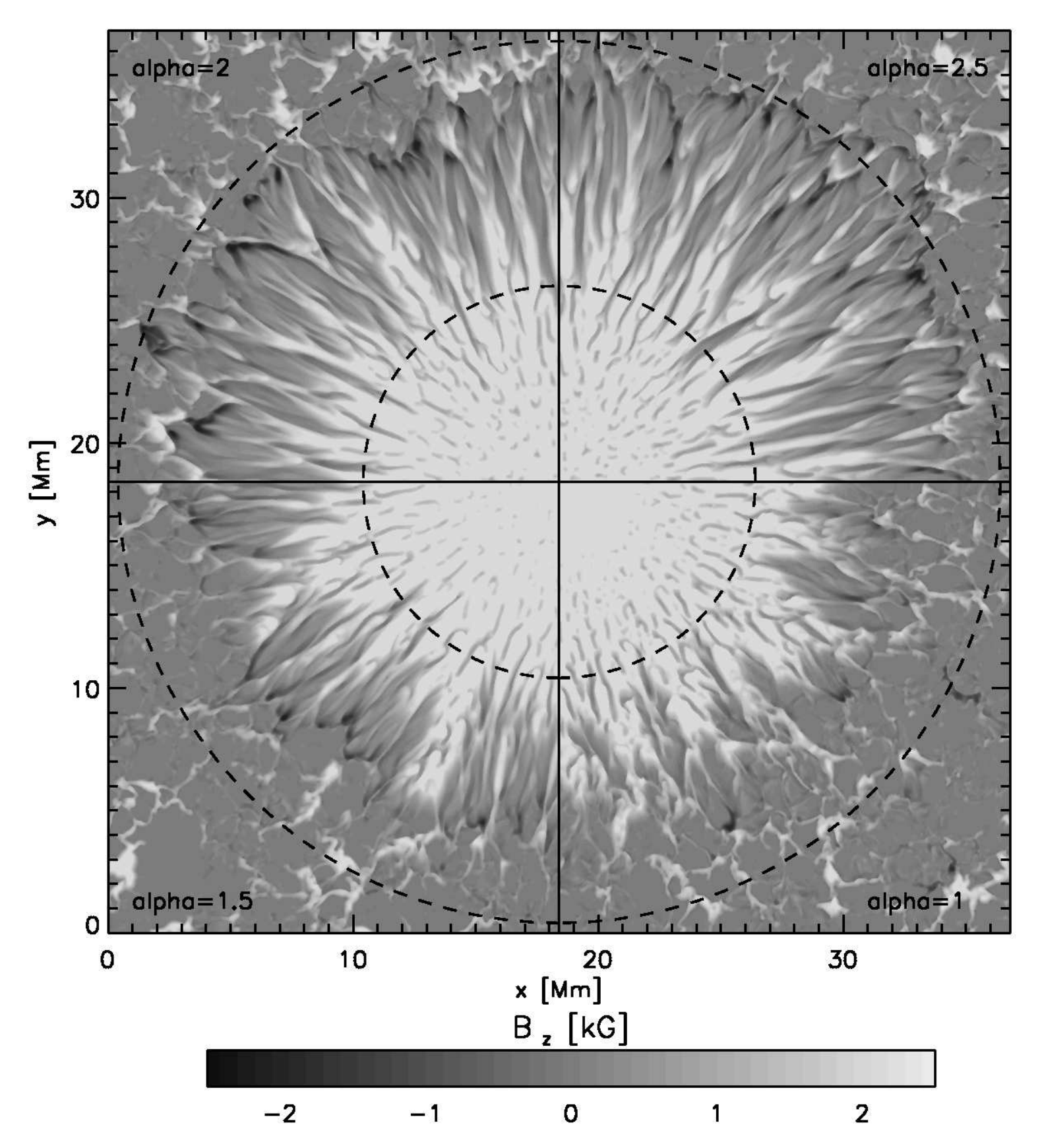}}
  \caption{Influence of the magnetic boundary condition at the top of
    the domain on the radial extent of the penumbra. Left: bolometric
    intensity, right: magnetogram at $\tau=1$. In each panel the 4 quadrants
    correspond to simulations performed with different magnetic top
    boundary conditions, (different values for $\alpha$ as described in 
    Appendix B). The choice of $\alpha=1$ corresponds to a potential field
    extrapolation. Note that all simulations were performed in a 49 Mm wide
    domain, we show here only subsections.
  }
  \label{fig:f1}
\end{figure*}
\begin{figure*}
  \centering 
  \resizebox{0.7\hsize}{!}{\includegraphics{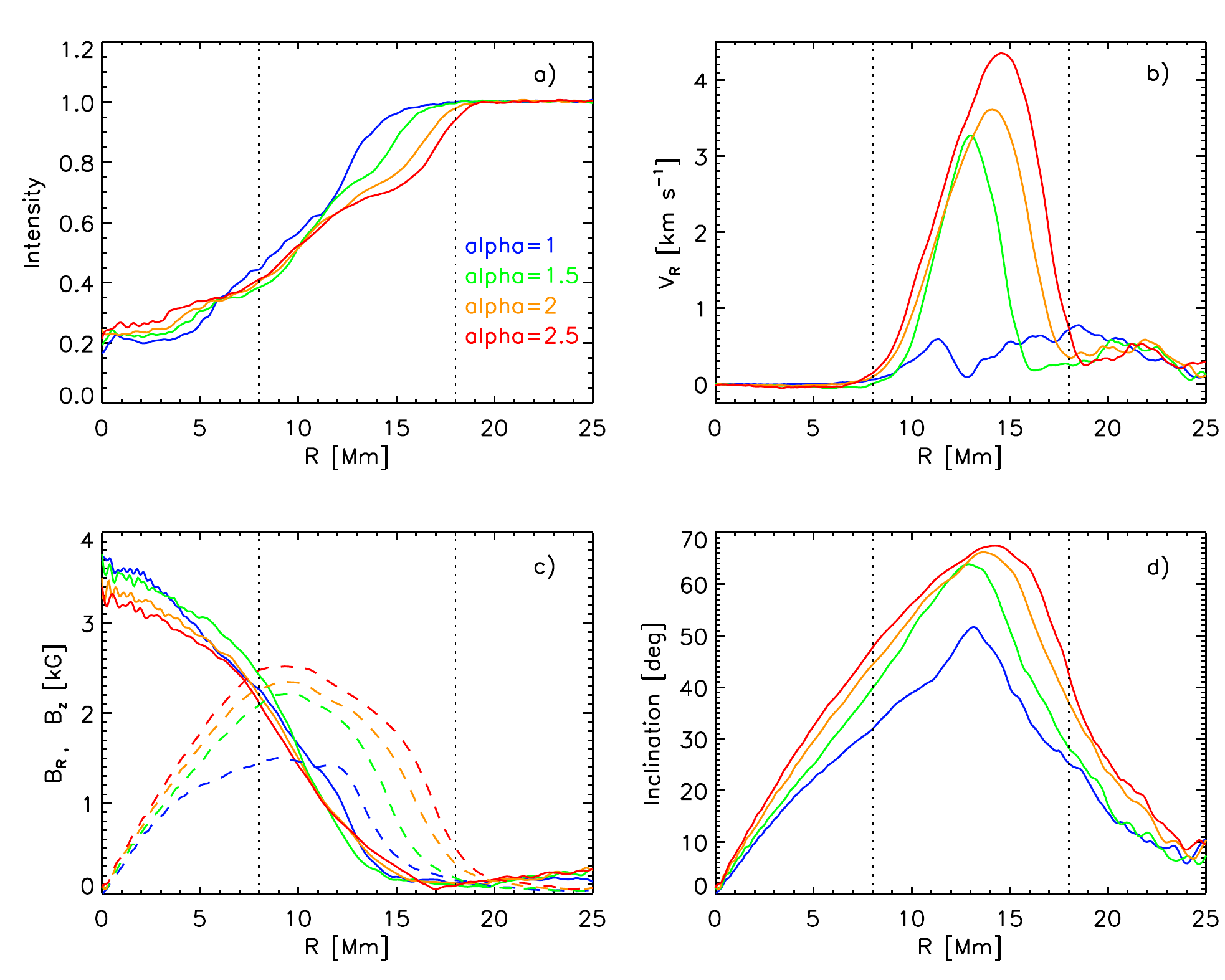}}
  \caption{Azimuthally averaged quantities at the $\tau=1$
    level for the different top boundary conditions shown in Figure
    \ref{fig:f1}: a) bolometric intensity, b) radial
    (Evershed) flow velocity, c) vertical (solid) and radial (dashed) field
    components, and d) field inclination angle. Increasing values for $\alpha$
    lead to more extended penumbrae with faster Evershed flows, the most 
    dramatic change is from $\alpha=1$ to $\alpha=1.5$. In terms of the 
    inclination angle the potential field boundary falls short of the other
    solutions by about 10-15 degrees.}
  \label{fig:f2}
\end{figure*}

\begin{figure*}
  \centering 
  \resizebox{0.95\hsize}{!}{\includegraphics{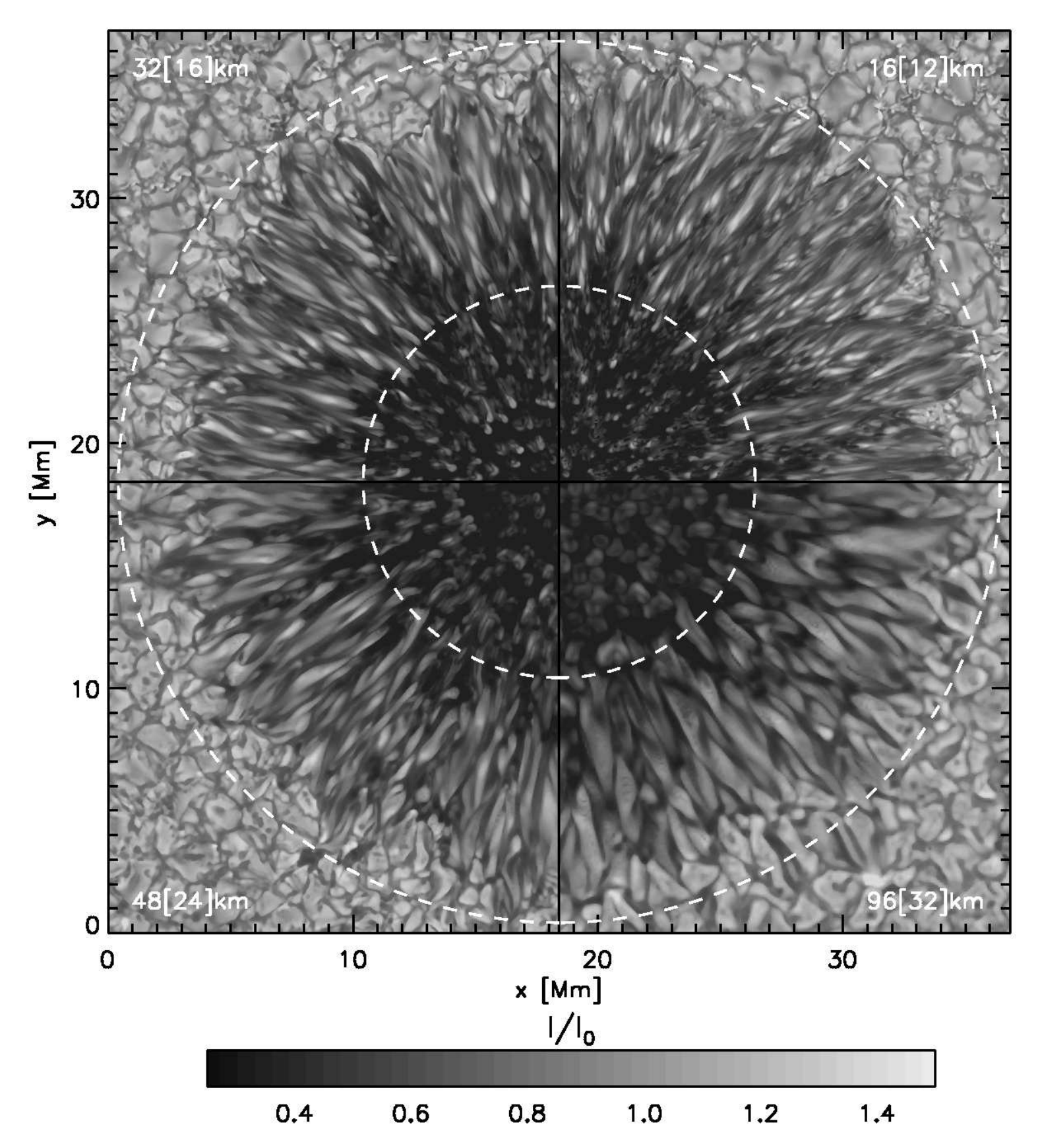}
    \includegraphics{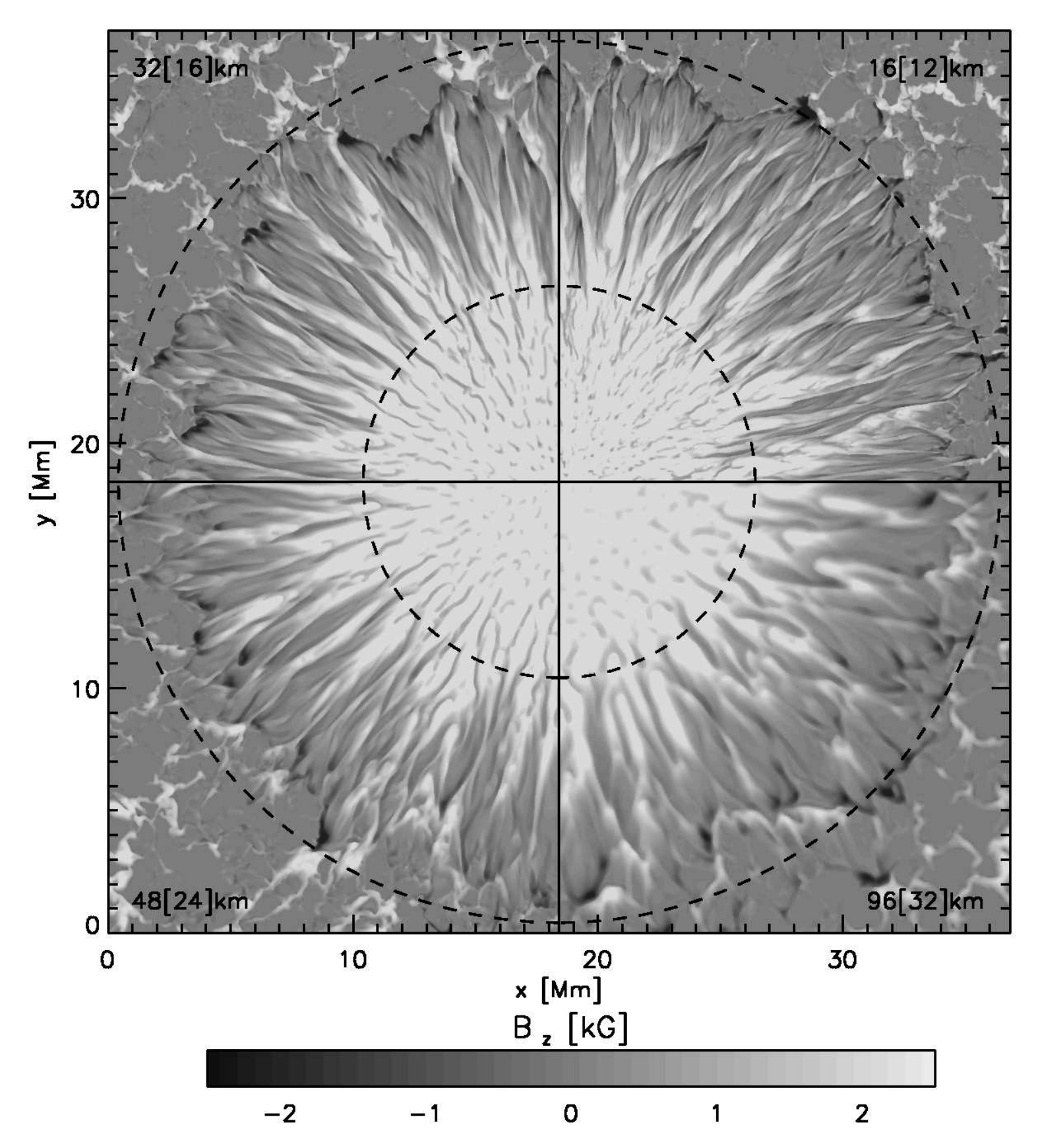}}
  \caption{Influence of the numerical grid resolution on the properties
    of the penumbra. Left: bolometric intensity, right: magnetogram at 
    $\tau=1$. In each panel the 4 quadrants correspond to simulations 
    performed with different resolution as indicated in the corners.
    Note that all simulations were performed in a 49 Mm wide
    domain, we show here only subsections.}
  \label{fig:f3}
\end{figure*}

\begin{figure*}
  \centering 
  \resizebox{0.7\hsize}{!}{\includegraphics{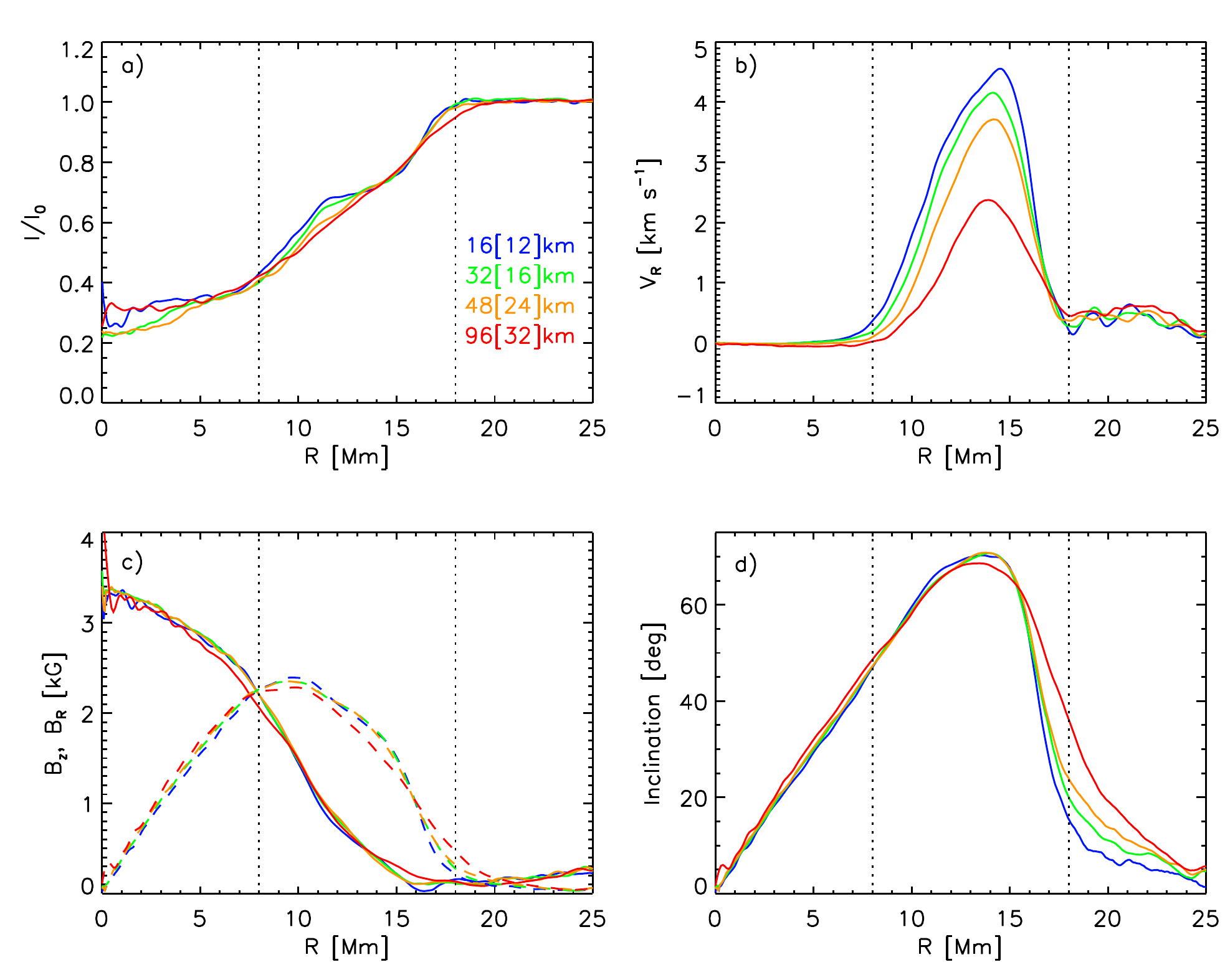}}
  \caption{Azimuthally averaged quantities at the $\tau=1$
    level for the different grid resolutions shown in Figure
    \ref{fig:f3}: a) bolometric intensity, b) radial
    (Evershed) flow velocity, c) vertical (solid) and radial (dashed) field
    components, and d) field inclination angle.
  }
  \label{fig:f4}
\end{figure*}

For the magnetic field we use a boundary condition that allows us to manipulate 
the inclination angle of the magnetic field. We focus on a 
single sunspot, which implies that the horizontal boundary imposes same 
polarity spots nearby and one thus produces only a very subdued penumbra
\citep{Rempel:etal:Science}. A relaxation of the horizontal periodicity is 
non-trivial and we thus decided to use a boundary that allows to control the 
inclination angle of the magnetic field to mimic different global field 
configurations, which is described in detail in Appendix B. A free parameter, 
$\alpha$, gives for $\alpha=1$ a potential field extrapolation subject
to horizontal periodicity, i.e. the field becomes vertical asymptotically.
Values of $\alpha>1$ lead to more horizontally inclined field, i.e. the
horizontal field component is approximately enhanced by a factor of $\alpha$
compared to the potential field reference. The resulting field 
extrapolations are for $\alpha\ne 1$ not force free outside the computational 
domain. They are not intended to give a realistic description for the global 
field topology above a sunspot, rather to explore the influence of different 
field geometries on the structure of a sunspot penumbra. 

All of the simulations presented here were computed with gray radiative 
transfer. 
To allow for a better comparison with observations through forward modeling
of spectral lines we computed 2 non-gray models that are derived from the
gray simulations. We have a non-gray version of the simulation
with $32\,[16]$ km grid resolution that was advanced for 26 minutes and we
have a non-gray model with $12\,[8]$ km resolution that was restarted from the
upper half of our $16\,[12]$ km case and advanced for 15 minutes. The non-gray
model with $32\,[16]$ km was analyzed by \citet{Bharti:etal:2011:conv},
more results will be presented in forthcoming publications.

In the following analysis we will focus entirely on the gray simulations
and discuss penumbral structure mostly through quantities extracted on constant 
$\tau$ surfaces. In Figure \ref{fig:f0} we display a snapshot from our highest
resolution gray simulation ($16\,[12]$ km) that we consider in this paper.
The simulation was computed with the $\alpha=2$ boundary condition and evolved 
for $1$ hour at the highest resolution.

\section{Global sunspot properties}
\label{sect:global}

\subsection{Influence from magnetic top boundary condition} 
In Figure \ref{fig:f1} we show 4 sunspot models computed with the boundary 
conditions $\alpha=1, 1.5, 2, 2.5$. The case $\alpha=1$ (lower right 
quadrant) leads to a very subdued penumbra with a few isolated filaments,
which is consistent with the penumbra structure that was found in 
\citet{Rempel:etal:Science} in the direction where the periodicity imposes
the same polarity sunspots. The other three cases show more developed
penumbrae, the overall radial extent increases with the value of $\alpha$. In
Figure \ref{fig:f2} we show azimuthal averages of (a) intensity, (b) 
Evershed flow, (c) vertical and radial magnetic field, and (d) inclination 
angle.
These quantities are in addition averaged for about 1 hour in time. 
Apart from the monotonic increase of the penumbra extent with $\alpha$, it is
also evident that the most dramatic differences occur between the the potential
field case ($\alpha=1$) and the other three cases with $\alpha>1$. For
$\alpha=1$ the azimuthally averaged radial flow velocity stays around 
$500$~m$\,$s$^{-1}$, while all other cases have outflows with more than 
$3$~km$\,$s$^{-1}$ on average. Similarly the curves for the radial magnetic
field as well as inclination angle do not differ as much between the 
$\alpha>1$ cases as compared to $\alpha=1$ and the rest. While the potential 
field case reaches only an average inclination of $50^{\circ}$, all the other 
solutions with more extended penumbra and Evershed flow reach about 
$65^{\circ}$.

This leads to the surprising conclusion that the potential field case (which
is the only physical, i.e. force free solution outside the computational 
domain) is a clear outlier compared to the rest. This result has to be seen
in the context of horizontal periodicity, which is not the proper horizontal
boundary condition to describe the magnetic field above sunspots. A more
reasonable (but computationally more difficult to implement) boundary
condition would be the free expansion of magnetic field into a half room
(perhaps considering in addition spherical geometry), which would lead naturally
to a magnetic field with stronger horizontal field components. The boundary
conditions with $\alpha>1$ emulate that behavior while maintaining horizontal
periodicity. Note that simply increasing the horizontal extent of the domain 
for $\alpha=1$ does not alleviate this problem, since the field remains
asymptotically vertical.
For the following investigation we will use a top boundary condition with 
$\alpha=2$.
    
\subsection{Influence from numerical grid resolution}
\label{sect:grid_res}
We modify the numerical grid resolution in the range from $96\,[32]$ to
$16\,[12]$~km (horizontal [vertical] resolution). The sunspot models computed 
with different grid resolutions are not fully independent from each other: 
The model with a resolution of $48\,[24]$~km resolution was started from a 
snapshot of the $96\,[32]$~km resolution simulation evolved for 1 hour and 
ran for an additional $5.5$ hours. The model with a resolution of $32\,[16]$~km
resolution was started from a snapshot of the $48\,[24]$~km resolution
run after 3.3 hours and ran for an additional 2.2 hours. The highest 
resolution case with $16\,[12]$~km resolution was started from the last 
snapshot of the $32\,[16]$~km case and evolved for an additional hour. 
Overall we compare sunspot models that have been evolved for about $6$
hours from the initial state. We did not evolve the higher resolution cases 
for the full length of time because of their computational expense (1 hour 
at $16\,[12]$~km resolution, $3072\times 3072\times 512$ grid points,
costs about 800,000 CPU hours on a CRAY XT-5). For the small length and 
associated short time scales in the photosphere about $1$ hour at the 
highest resolution is sufficient to allow for the solution to adapt to 
the grid spacing. Structures in the deeper parts of the domain are already 
well resolved at lower resolution. 

\begin{figure*}
  \centering 
  \resizebox{0.95\hsize}{!}{\includegraphics{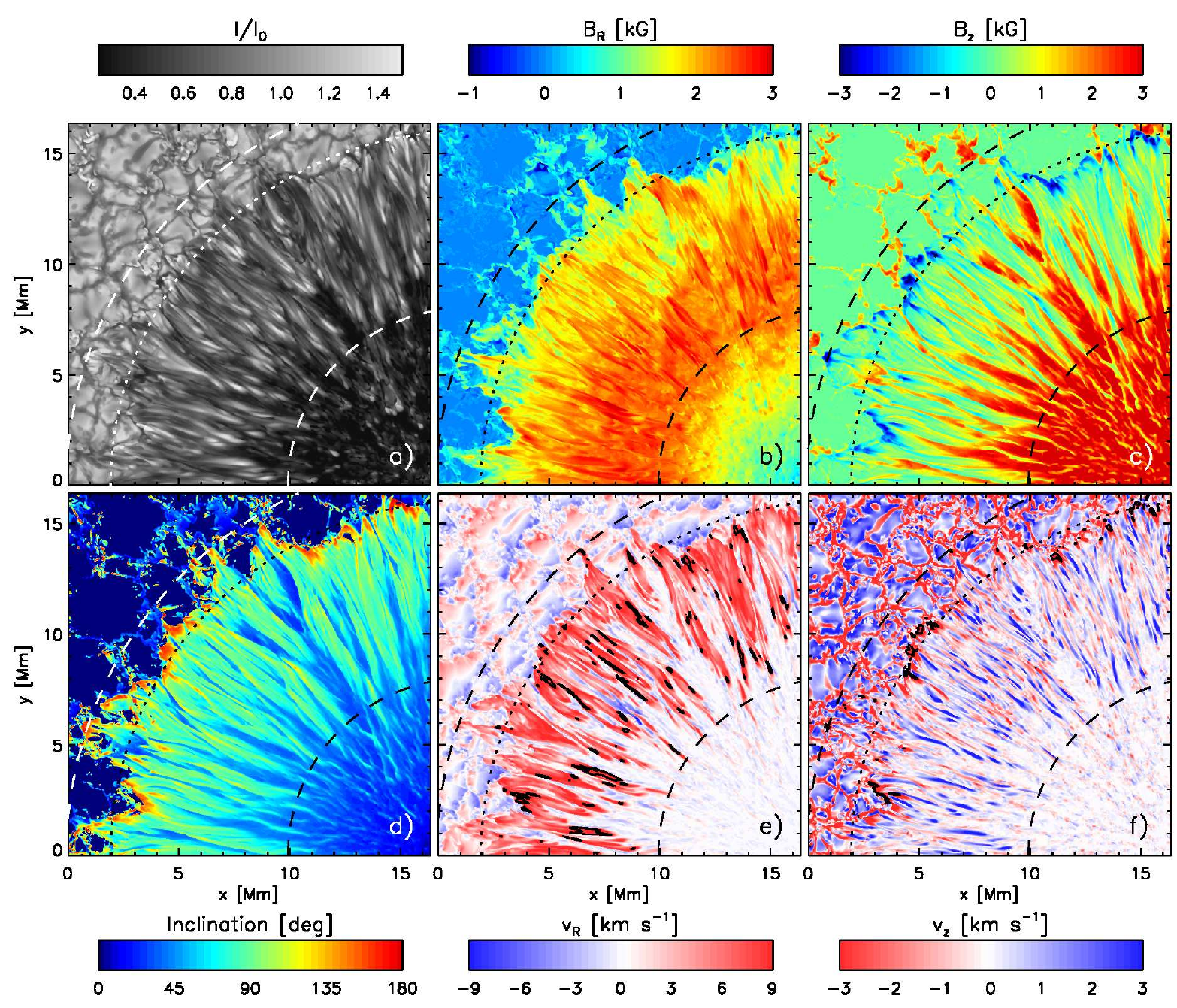}}
  \caption{Sunspot fine structure at $\tau=1$ for the highest resolution
    run (16 [12]~km) with the $\alpha=2$ top boundary condition: a) bolometric
    intensity, b) radial field strength, c) vertical field strength, d) 
    inclination angle, e) radial (Evershed) flow velocity, and f) vertical
    velocity. Black contours indicate in panel e) outflows with more than 
    10~km$\,$s$^{-1}$, and in panel f) supersonic downflows. The latter coincide
    mostly with strong inverse polarity patches in panel c). The dashed
    circles indicate R=8 and R=18 Mm, the dotted circle R=16 Mm. An animation
    is provided with the online version.}
  \label{fig:f5}
\end{figure*}

\begin{figure*}
  \centering 
  \resizebox{0.7\hsize}{!}{\includegraphics{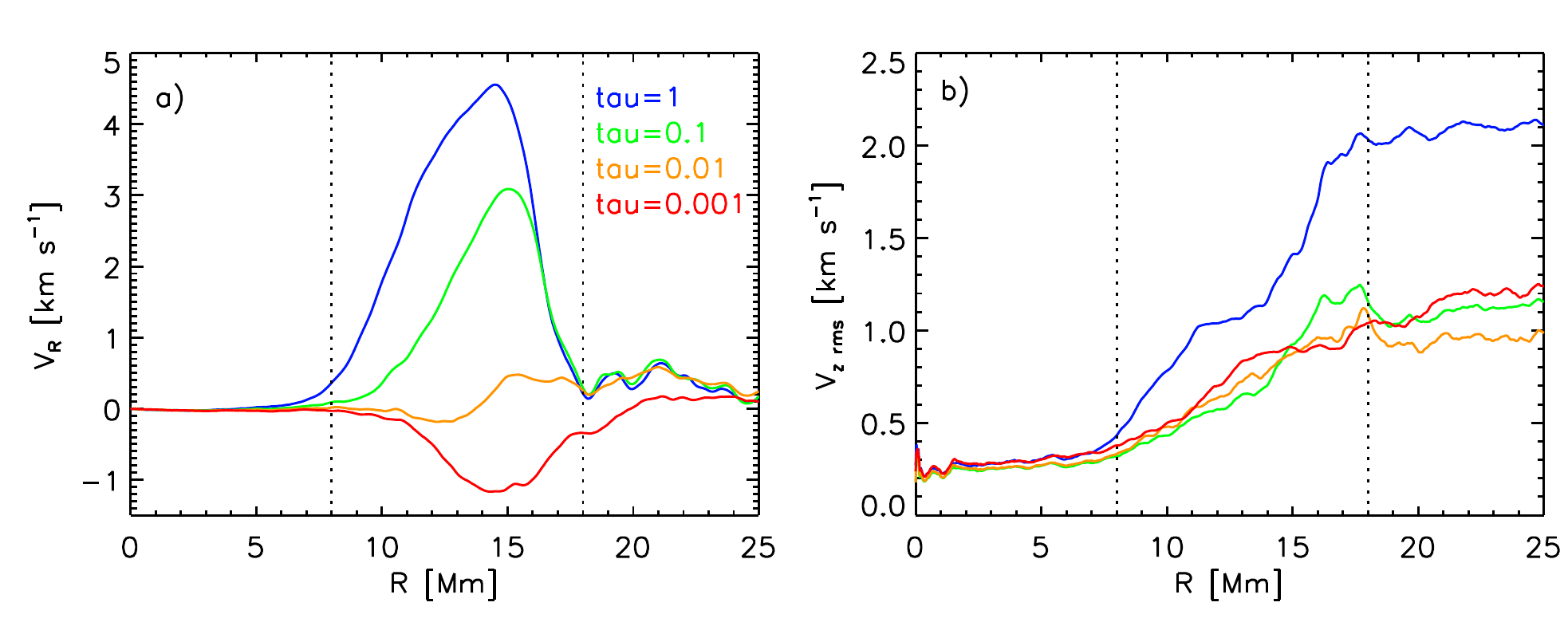}}
  \caption{Height dependence of a) azimuthally averaged Evershed flow and
    b) vertical rms velocity.}
  \label{fig:f6}
\end{figure*}

\begin{figure*}
  \centering 
  \resizebox{0.68\hsize}{!}{\includegraphics{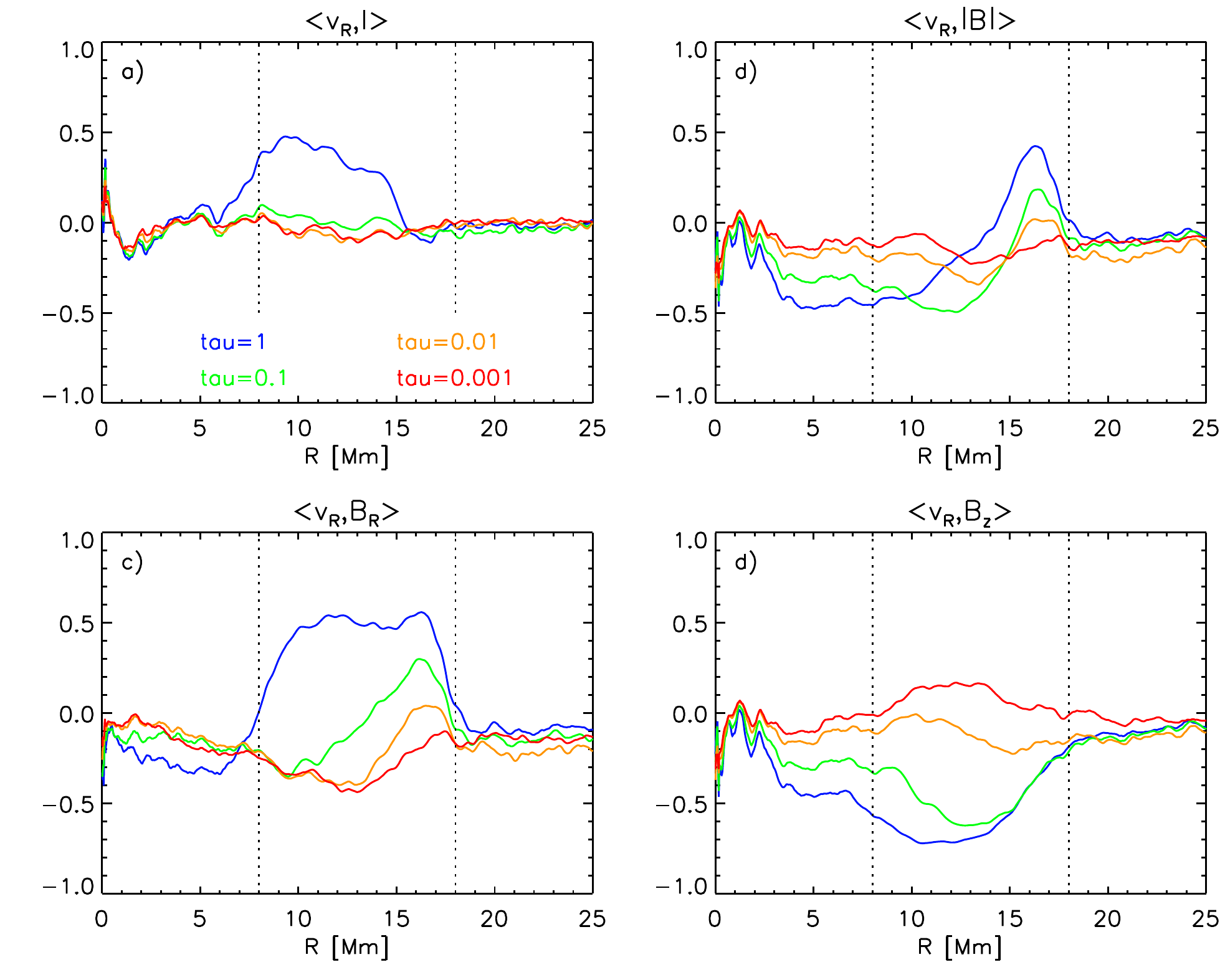}}
  \caption{Correlations of several quantities displaying penumbral fine
    structure as function of height: a) $v_R-I$, b) $v_R-|B|$, c) $v_R-B_R$, 
    and d) $v_R-B_z$.}
  \label{fig:f7}
\end{figure*}

\begin{figure*}
  \centering 
  \resizebox{0.68\hsize}{!}{\includegraphics{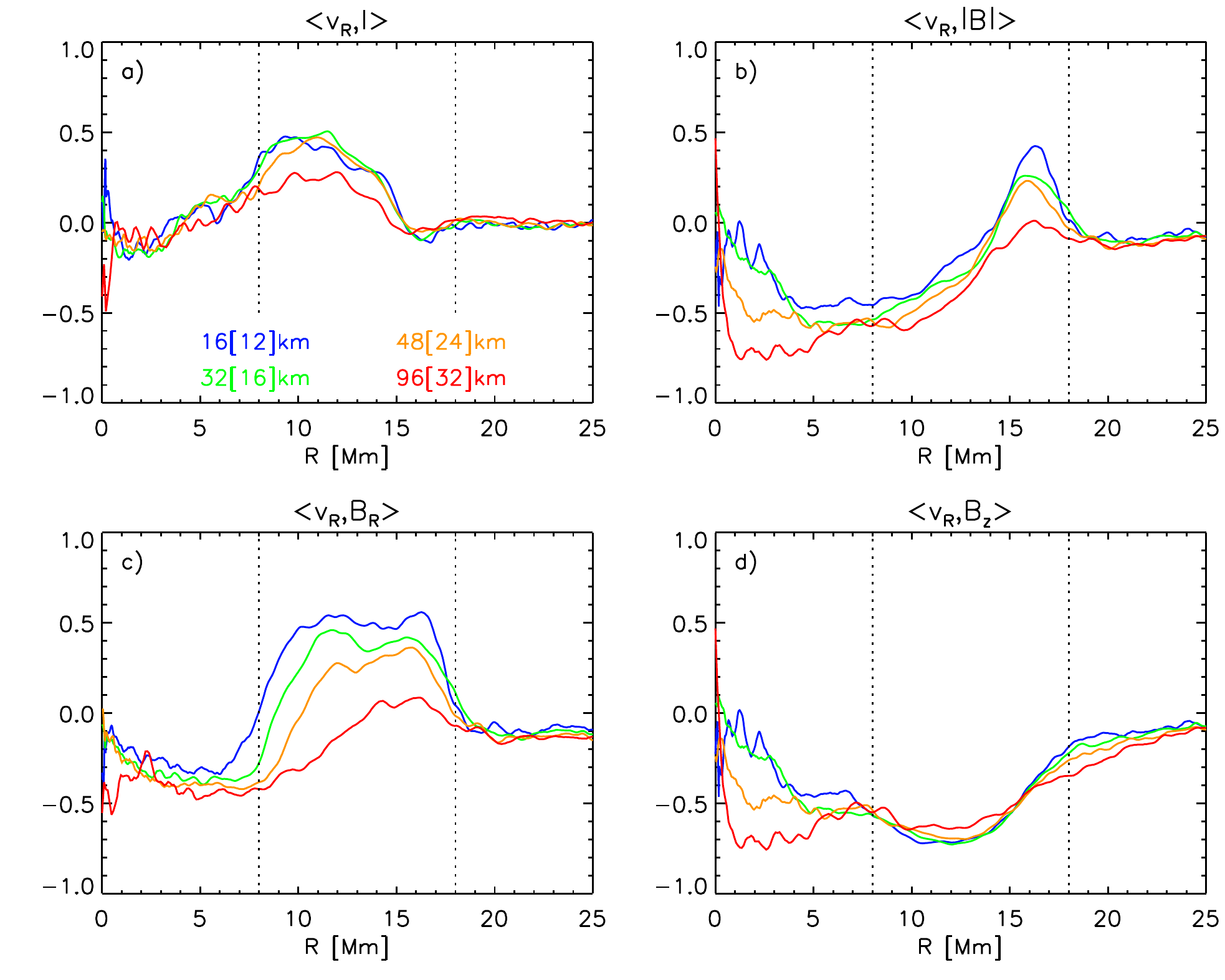}}
  \caption{Same correlations as shown in Figure \ref{fig:f7}. All
    quantities are evaluated at the $\tau=1$ for different grid 
    resolutions (color).}
  \label{fig:f8}
\end{figure*}

Figure \ref{fig:f3} shows the (gray) intensity and magnetogram at $\tau=1$; 
Figure \ref{fig:f4} shows the corresponding 
azimuthal averages of quantities at $\tau=1$. Overall there is no significant 
influence from grid resolution on the radial extent of the penumbra as well
as the magnetic structure at $\tau=1$, displayed in panels 
(c) and (d). Significant differences occur with regard to the average
intensity profile and Evershed flow shown in panels 
(a) and (b). Only the higher resolution cases show the formation of a plateau 
like feature at around $0.7 I_{\odot}$ in the inner penumbra. The average 
Evershed flow speed is increasing monotonically with resolution from
a peak at around $2.3$~km$\,$s$^{-1}$ for $96\,[32]$~km resolution to about
$4.6$~km$\,$s$^{-1}$ for $16\,[12]$~km resolution. However, we see also here
a clear sign of convergence: while the relative resolution changes from 
$96\,[32]$ to $48\,[24]$~km and from $32\,[16]$ to $16\,[12]$~km are the same,
the relative changes in the peak flow velocities are $1.55$ and $1.1$, 
respectively. The strong dependence of the Evershed flow velocity on grid
resolution is a consequence from the driving mechanism that is concentrated in
a thin boundary layer just beneath $\tau=1$ \citep{Rempel:2011}, which is 
difficult to resolve numerically (see also Section \ref{sect:subsurf}). 
The umbra is with $0.3 I_{\odot}$ rather bright for a sunspot with an
umbral field strength exceeding $3$~kG. We return to this point in Section 
\ref{sect:comment_diff}.   


\section{Photospheric fine structure}
\label{sect:finestruct}
Figure \ref{fig:f5} presents the penumbral fine structure at the $\tau=1$
level for our highest resolution case ($16\,[12]$~km resolution). This figure is
very similar to Figure 3 in \citet{Rempel:2011}, in which we displayed the
fine structure at $32\,[16]$~km resolution for the double sunspot simulation of 
\citet{Rempel:etal:Science}. The large degree of similarity underlines that
the details of fine structure are not very sensitive to the numerical setup.
Penumbral
filaments show an enhancement of the radial magnetic field component (panel b),
while the vertical field component is strongly reduced (panel c). In the outer 
penumbra we find a substantial amount of opposite polarity flux, which will be
characterized further in Section \ref{sect:polarity}. The combination of a
enhanced radial magnetic field component with a reduced vertical field 
component results in the uncombed structure of the penumbra
with a strong variation of the field inclination angle (panel d). Fast
horizontal (Evershed) outflows are found along the almost horizontal stretches 
of magnetic field (panel e). Overturning convection (panel f) is the underlying
driving mechanism, which will be characterized further in Section 
\ref{sect:conv}. We provide an animation of Figure \ref{fig:f5} with the
online version.

Figure \ref{fig:f6} presents the height dependence of the azimuthally averaged
Evershed flow and the vertical rms velocity. As already
pointed out in \citet{Rempel:2011}, the Evershed flow peaks in the deep
photosphere and falls off rapidly with height; we see the transition to an
inverse Evershed flow at about $\tau=0.01$. This figure is qualitatively
very similar to the observations reported by 
\citet[][see Figure 8 therein]{Bellot-Rubio:etal:2006} and by 
\citet{Borrero:etal:2008}. In Figure \ref{fig:f6}b we show
the height dependence of the vertical rms velocity. We find essentially a 
drop by a factor of $2$ between the $\tau=1$ and $\tau=0.1$ levels, but
no further drop toward higher layers. 

\begin{figure*}
  \centering 
  \resizebox{0.85\hsize}{!}{\includegraphics{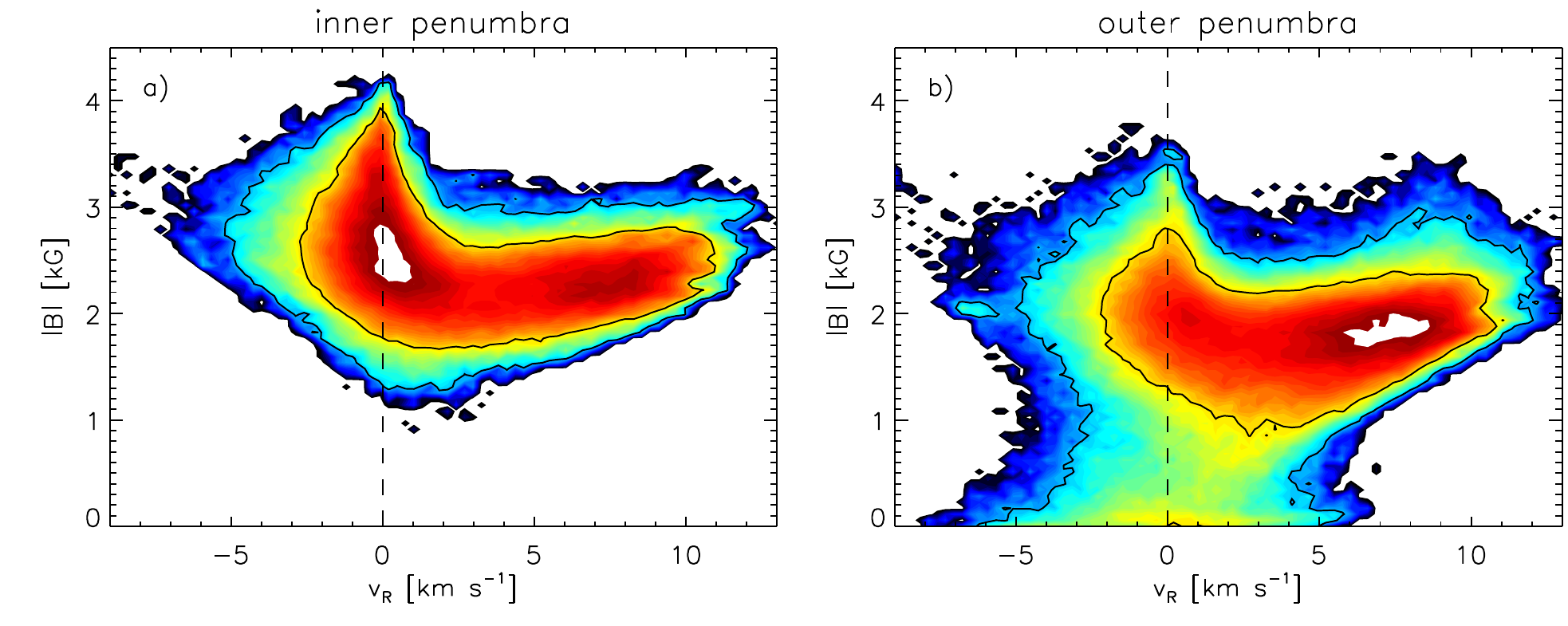}}
  \caption{Bivariate probability density function for $v_R$ and $\vert B\vert$
    at  $\tau=1$ for the highest resolution case. Panel a) shows the inner
    penumbra ($R=10$ to $13$~Mm), panel b) shows the outer penumbra
    ($R=13$ to $16$~Mm).}
  \label{fig:f9}
\end{figure*}

\subsection{Robustness of penumbral fine structure}
To characterize fine structure further and quantify the resolution dependence
of quantities we present in Figure \ref{fig:f7} correlations between the 
Evershed flow velocity and (a) intensity, (b) magnetic field strength as well 
as (c) radial and (d) vertical magnetic field components. All correlations 
are computed based on fluctuations around a smooth background that was
obtained through a convolution with a Gaussian having a FWHM of $1.5$~Mm.
At $\tau=1$
the Evershed flow is found in the inner penumbra in bright features, whereas
the correlation vanishes toward the outer penumbra (panel a). A similar 
behavior was also found by \citet{Schlichenmaier:etal:2005} and 
\citet{Ichimoto:etal:2007}. The correlation between Evershed flow and 
magnetic field strength is negative in the inner half, but positive in the
outer half of the penumbra. An Evershed flow in stronger magnetic field
regions in the outer penumbra was also inferred by \citet{Tritschler:etal:2007}
and \citet{Ichimoto:etal:2008} from net circular polarization (NCP) variations
with the viewing angle. The sign change in the $v_R-\vert B\vert$ correlation
results from two effects: a positive correlation with the radial field 
component and a negative correlation with the vertical field component 
throughout the penumbra (Figure \ref{fig:f7}c,d). While the negative 
correlation 
with the vertical field dominates the inner penumbra, the positive correlation
with the radial field component dominates the outer penumbra (see also Figure 
\ref{fig:f5}). Except for the correlation between $v_R$ and $B_z$ that is with
values about $-0.7$ quite significant all others reach in the deep photosphere 
only moderate levels of up to $0.5$. All correlations decay very rapidly with 
height.

To quantify the robustness of photospheric fine structure we focus now on the
correlations at $\tau=1$ and study the resolution dependence in Figure 
\ref{fig:f8}. The lowest resolution case ($96\,[32]$~km) somewhat misses 
the $v_R-I$ correlation in the inner and the $v_R-\vert B\vert$ correlation 
in the outer penumbra, all other grid resolutions produce comparable levels
of correlations. We can conclude that the magneto-convection process 
underlying penumbral fine structure is well captured starting from a 
resolution of about $48\,[24]$~km. This resolution also produces an average 
Evershed flow amplitude at this resolution that differs less
than $20\%$ from our highest resolution case. As we will discuss in Section
\ref{sect:subsurf}, the trend of increasing magnetic field in flow channels 
(see $v_R-B_R$ correlation) and the resolution dependence of the Evershed flow 
have a common origin.

\subsection{Magnetized flow channels}
\label{sect:Bevershed}
The positive $v_R-B_R$ correlation throughout the penumbra increases in
amplitude with increasing resolution, which strongly points toward an 
active process
supporting strong horizontal field in the flow channels as opposed to remnant
magnetic field due to unavoidable numerical diffusion effects as speculated
by \citet{Nordlund:Scharmer:2010}. It was pointed out by \citet{Rempel:2011}
that the horizontal field originates mostly from the horizontal shear in the 
sub-photospheric Evershed flow, i.e. the term $B_z\partial_zv_R$ in the 
induction equation (see Section \ref{sect:subsurf}). The robustness of this 
effect points toward a strongly magnetized Evershed flow in photospheric 
layers as it has been inferred from spectropolarimetric observations 
\citep[see, e.g.][]{Ichimoto:etal:2008,Borrero:Solanki:2008,
Borrero:Solanki:2010,Borrero:2009}. This result is not necessarily in 
contradiction to the 'gappy' penumbra
model of \citet{Spruit:Scharmer:2006} and \citet{Scharmer:Spruit:2006} since 
the strong horizontal magnetic field is confined to the thin boundary layer at
$\tau=1$ \citep{Rempel:2011}. The value of $\vert B_R\vert$ is in flow channels
typically about a few 100~G stronger than in the background. In Figure
\ref{fig:f9} we present bivariate probability density function for $v_R$ and
$\vert B\vert$ in the inner penumbra ($R=10$ to $13$~Mm) in panel a) and the 
outer penumbra ($R=13$ to $16$~Mm) in panel b). In the inner penumbra we find
field strength of up to $4$~kG in regions with low radial velocity, which
correspond to the spines. Fast outflows have field strength of more than 
$2$~kG, mostly due to $B_R$. We see a similar behavior in the outer penumbra
with reduced overall field strength and less pronounced spines, but also
here fast outflows are associated with field strengths around $2$~kG.
The extension toward low field strength is caused by a few granules
present in the region $13~\mbox{Mm} < R < 16~\mbox{Mm}$. In both, inner and outer penumbra, there is also a population of inflows with 
substantial field strength. We see also a trend for increasing $\vert B\vert$
with increasing outflow velocities ($v_R>0$).

\subsection{Inverse polarity flux}
\label{sect:polarity} 

\begin{figure*}
  \centering 
  \resizebox{0.7\hsize}{!}{\includegraphics{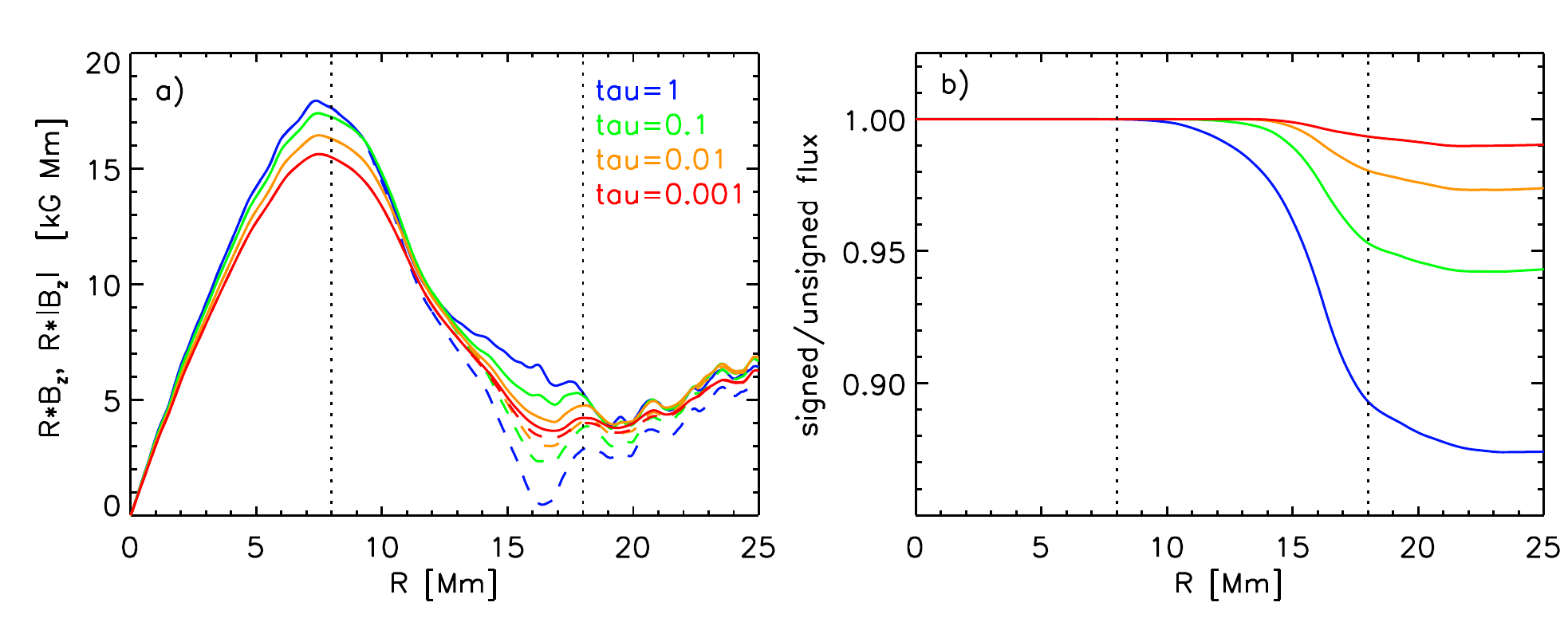}}
  \caption{Quantification of inverse polarity flux in the penumbra for 
    different height levels. Panel a) shows the quantities 
    $R \langle B_z\rangle$ (dashed)  and $R \langle |B_z| \rangle$ (solid)
    as function of radius (compare to Figure 13 in
    \citet{Sanchez-Almeida:2005:sunspot}). Panel b) shows the ratio of the 
    radially integrated signed and unsigned fluxes.}
  \label{fig:f10}
\end{figure*}

\begin{figure*}
  \centering 
  \resizebox{0.7\hsize}{!}{\includegraphics{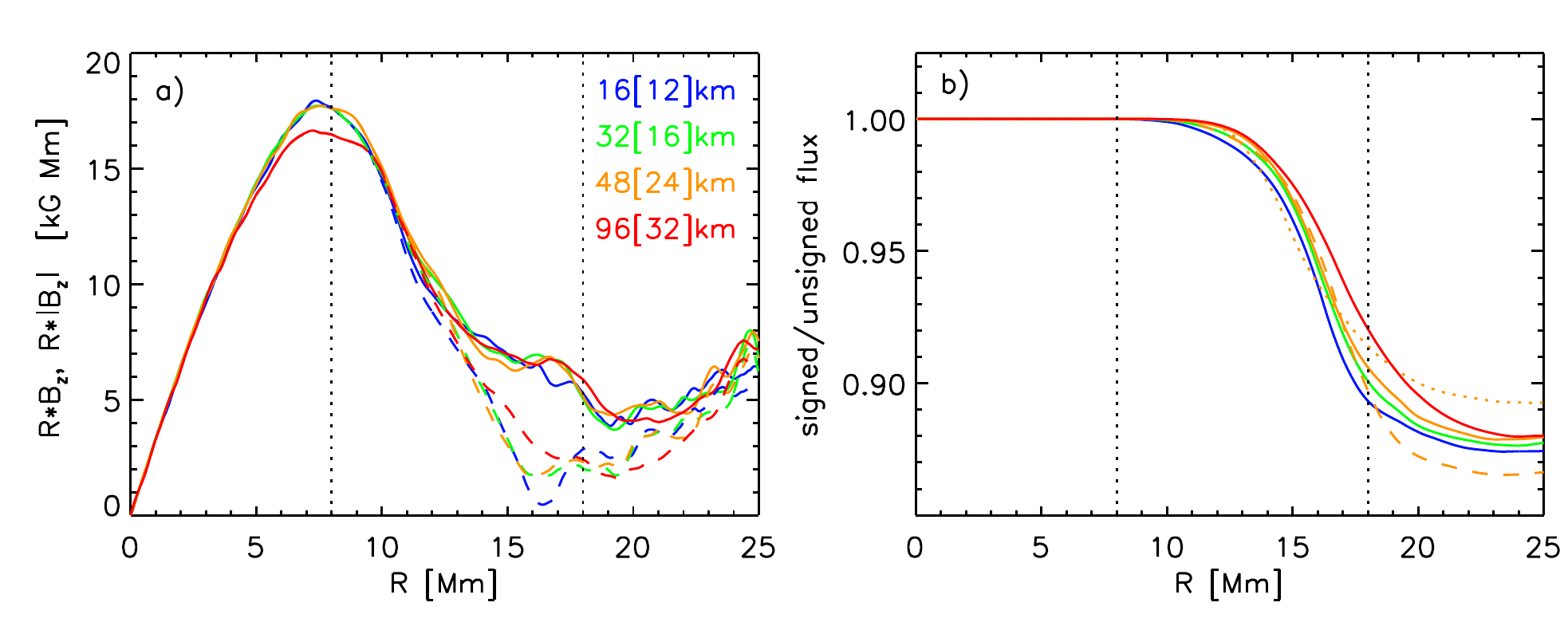}}
  \caption{Quantification of inverse polarity flux in the penumbra at 
    $\tau=1$ for different grid resolutions (color). Panel a) shows the 
    quantities $R \langle B_z\rangle$ (dashed) and $R \langle |B_z| \rangle$ 
    (solid) as function of radius, panel b) the ratio of the radially 
    integrated signed and unsigned fluxes. In panel b) the dotted (dashed)
    lines show for comparison results with different top boundary conditions
    $\alpha=1.5 (2.5)$ for the $48\,[24]$~km case.
    }
  \label{fig:f11}
\end{figure*}

Inverse polarity magnetic flux embedded in the sunspot penumbra is an
integral part of penumbral fine structure. Due to the substantial amount
of overturning convection (see Sect. \ref{sect:conv}) a certain amount
of magnetic field lines has to turn back into the photosphere within
the penumbra. Return flux along the outer penumbra rim was first found by
\citet{Westendorp:etal:1997} and interpreted in the context of predictions
from flux tube models such as \citet{Thomas:Montesinos:1993} and
\citet{Montesinos:Thomas:1997}. Fast (supersonic) downflows and return flux in
the interior of the penumbra were found by \citet{DelToro:etal:2001}.
Substantial amounts of inverse polarity flux were deduced from observations
by \citet{Sanchez-Almeida:2005:sunspot}, which we will use as a reference for
the analysis presented here. This subject is not without controversy, for
example \citet{Langhans:etal:2005} did not find much evidence in
magnetograms at high spatial resolution.

In Figs. \ref{fig:f10} and \ref{fig:f11} we quantify the amount of inverse
polarity flux found in the penumbra. Following 
\citet{Sanchez-Almeida:2005:sunspot},
we show in the left panels of both figures the quantities 
$R \langle\vert B_z\vert\rangle$ (solid) and $R \langle B_z\rangle$ (dashed) as 
function of radius, where $\langle\ldots\rangle$ indicates the azimuthal
average. The right panels show the ratio of the radially integrated
signed and unsigned magnetic flux. Figure \ref{fig:f10} shows how these
quantities change with optical depth in the range from $\tau=1$ to
$\tau=0.001$. Significant amounts of inverse polarity flux are only present
in the deep photosphere, where we find integrated over the whole sunspot up
to $11\%$, already at $\tau=0.1$ only half of that remains. Figure \ref{fig:f11}
shows how these quantities vary with grid resolution as well as boundary
conditions. Comparing the integrated values at $R=18$ Mm in panel (b), we
see an increase of the inverse polarity flux from abut $8\%$ to $11\%$ over
the resolution range considered here (about half of the spread is due to our
lowest resolution run). We also find that more extended penumbrae tend to have
more opposite polarity flux than less extended ones (yellow dashed/dotted 
lines). Given the fact that the effective extent of the penumbra computed with
$\alpha=1.5\,(2.5)$ is $16\,(19)$~Mm, the corresponding fractions of opposite
polarity fluxes are $6\,(12)\%$. For a better comparison 
with the results from \citet{Sanchez-Almeida:2005:sunspot}, we present in
Table \ref{tab:t1} quantities that are comparable to those in their Table 1. 
Here the quantity $\Phi_z$ is defined analogous to theirs as
\begin{equation}
  \Phi_z(R_1<R<R_2)=2\pi\int_{R_1}^{R_2}R \langle B_z\rangle \mbox{d}R\;.
  \label{eq:e1}
\end{equation}
For computing the values in Table \ref{tab:t1} we used $R_s=18$ Mm for
the radius of the sunspot. We present here values computed for the highest 
resolution case. Our simulated sunspot is about a factor of
$1.6$ larger in terms of unsigned and a factor of $2$ larger in terms of
signed flux. While we find a large degree of qualitative agreement with 
\citet{Sanchez-Almeida:2005:sunspot}, the overall amount of inverse polarity 
flux falls short by a factor of about $3$ (first line in table).
We find a substantial amount of opposite polarity flux in the region
$12.7\,\mbox{Mm} < R < 18\,\mbox{Mm}$, where the signed flux is $50\%$ of the
unsigned flux. In contrast to  \citet{Sanchez-Almeida:2005:sunspot} we do not 
find opposite polarity flux at photospheric levels in the umbra (although, 
some amount of opposite polarity field is present in umbral dots beneath the 
$\tau=1$ surface in our simulations).

\begin{table*}
  \caption{Quantification of inverse polarity flux}
  \begin{center}
    \begin{tabular}{ l  l  l l }
      \hline
      \hline
      Location & Description & Unsigned & Signed \\
      \hline
      $\Phi_z(0<R<R_s)$ & full spot & $\Phi_0=1.18\times 10^{22}\mbox{Mx}$ 
      & $0.89\,\Phi_0$ \\
      $\Phi_z(R_s/2<R<R_s)$ & penumbra & $\Phi_1=0.44\,\Phi_0$ 
      & $0.76\,\Phi_1$ \\
      $\Phi_z(R_s/\sqrt{2}<R<R_s)$ & outer penumbra & $\Phi_2=0.19\,\Phi_0$ 
      & $0.49\,\Phi_2$ \\
      $\Phi_z(0<R<R_s/2)$ & umbra & $\Phi_3=0.54\,\Phi_0$ 
      & $1.00\,\Phi_3$ \\
      \hline
      \hline
      \label{tab:t1}
    \end{tabular}
  \end{center}
\end{table*}

\begin{figure*}
  \centering 
  \resizebox{0.7\hsize}{!}{\includegraphics{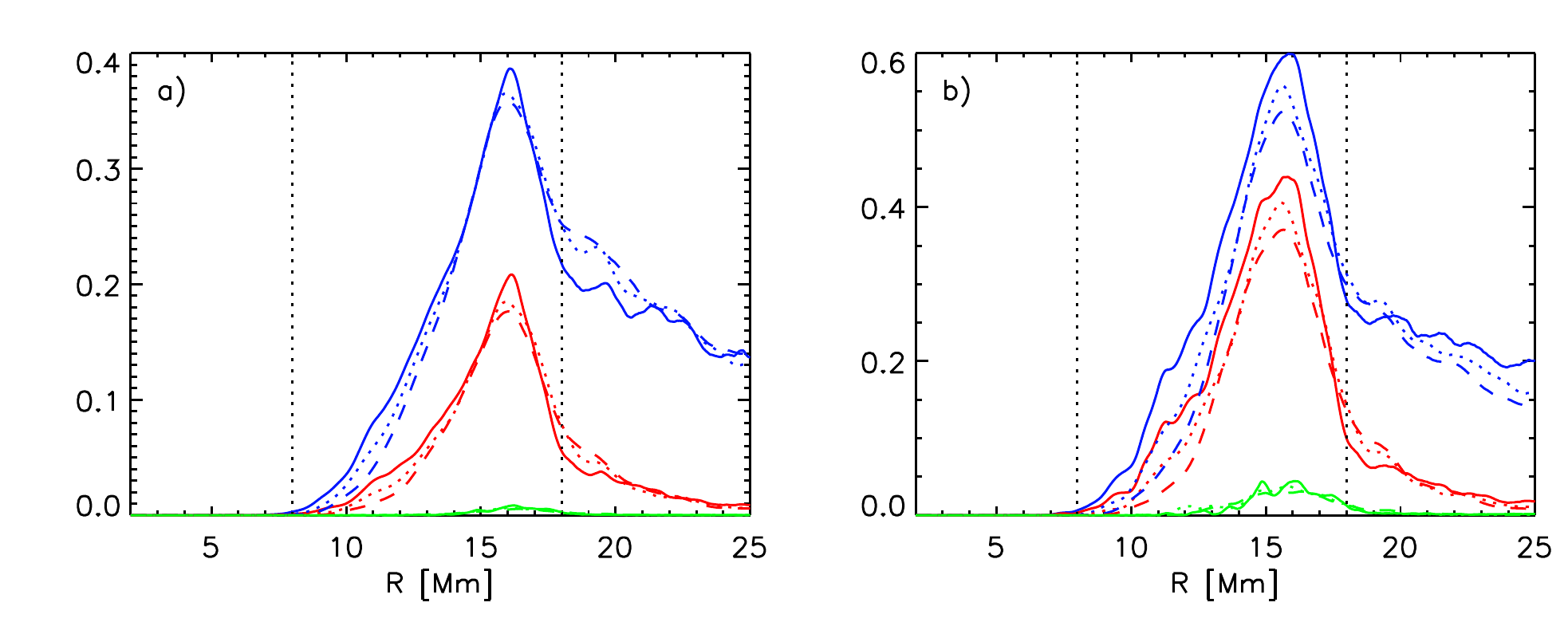}}
  \caption{a) Filling factors of regions with $B_z<-25$ G (blue),
    $B_z<-250$ G (red) and supersonic downflows (green) as function of 
    radius at $\tau=1$. b) Fraction of downward directed mass flux found 
    in these 
    regions as function of radius. We show quantities computed for the 
    resolutions of $16\,[12]$~km (solid), $32\,[16]$~km (dotted), and
    $48\,[24]$~km (dashed).
    }
  \label{fig:f12}
\end{figure*}
 
The fact that the amount of return flux in simulations shows only little 
resolution dependence might come as a surprise, since most of this flux is
present at small scales which vary somewhat with grid resolution. However,
return flux is a consequence of vigorous overturning convection,
which itself shows only little resolution dependence (see Section 
\ref{sect:conv}). Overturning convection is a direct consequence of energy 
flux constraints that are the same in all cases considered regardless of
resolution.

\subsection{Relation between opposite polarity regions and downflows}
\label{sect:vzbz_rel}
Figure \ref{fig:f12}a) presents the filling factors of regions with
opposite polarity field $B_z<-25$ G (blue) and $B_z<-250$ G (red). Both
peak at about $R=16$ Mm, which has been also indicated in Figure \ref{fig:f5}
through the dotted circle. The peak filling factors are $0.4$ and $0.2$ for
these two thresholds. The filling factor of supersonic downflows (green)
reaches about $1\%$. In Figure \ref{fig:f12}b) we evaluate the downward
directed mass flux in these regions relative to the total downward directed
mass flux as function of radius. Up to $60\%$ ($44$\%) of the returning 
mass flux is 
found in regions with $B_z<-25$ G ($B_z<-250$ G), up to $5\%$ in supersonic
downflows. Integrated over the entire penumbra (from $R=8$ to $R=18$ Mm)
about $40\%$ ($27\%$) of the downward directed mass flux is found in regions 
with $B_z<-25$ ($B_z<-250$). Supersonic downflows contribute only $2-3\%$.
Solid, dotted and dashed lines indicate the respective quantities for 
the resolution levels of $16\,[12]$, $32\,[16]$, and $48\,[24]$, respectively.

At $\tau=1$ the average velocity in the supersonic downflow regions is 
$-9.6$~km$\,$s$^{-1}$, with fastest flows reaching $-15$~km$\,$s$^{-1}$. The 
average 
magnetic field strength is $2.7$~kG. The average gas pressure is with 
$6.8\cdot 10^4$~dyne~cm$^{-2}$ about $30\%$ lower than the typical photospheric
pressure (at a constant height level the difference would be even larger), 
leading to an average intensity that is with $1.1 I_{\odot}$ clearly 
enhanced, in particular for a strong downflow. This has consequences for
the $I-v_z$ correlation in the outer penumbra, which we will discuss
further in Sections \ref{sect:conv} and \ref{sect:statistics} 

\begin{figure*}
  \centering 
  \resizebox{0.7\hsize}{!}{\includegraphics{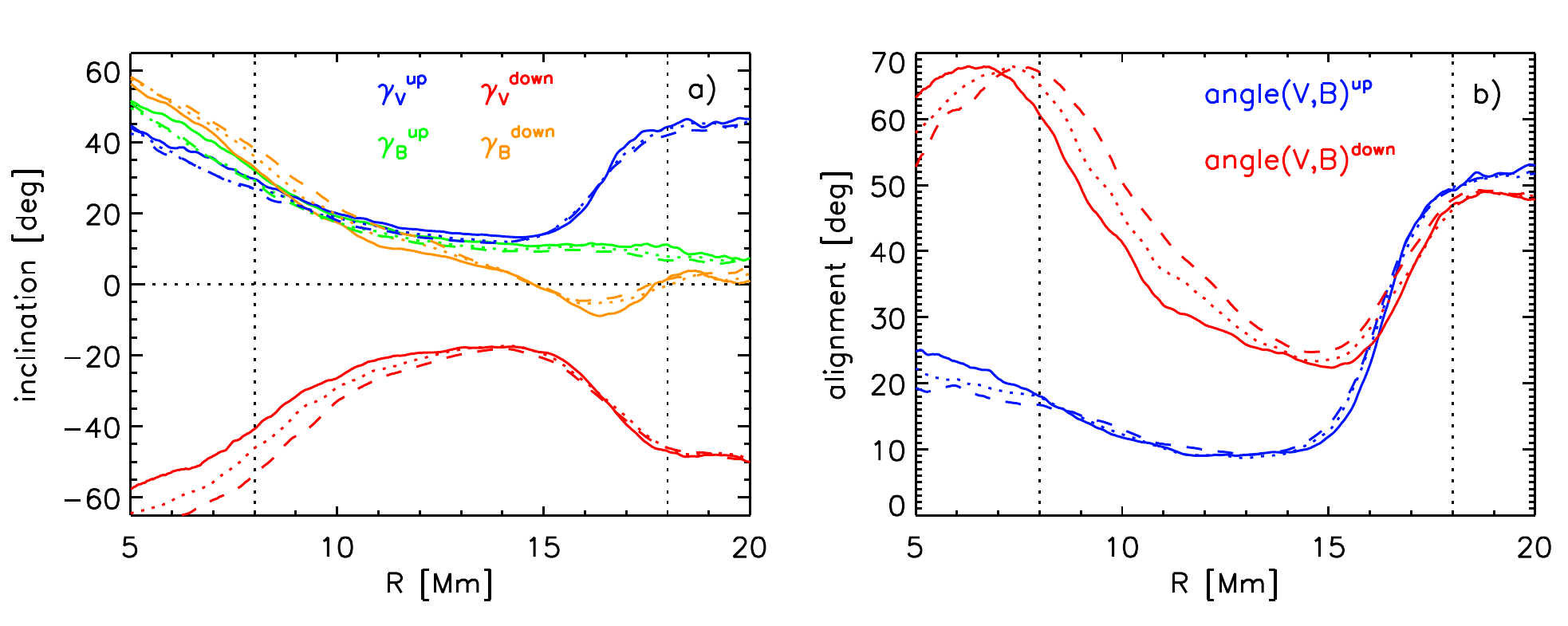}}
  \caption{a) Inclination of magnetic field and flow with respect to
    radial direction in Evershed flow channels at $\tau=1$, separated
    by upflow and downflow regions. b) Angle between $\vec{v}$ and
    $\vec{B}$ for the same regions. Different line styles indicate
    results from different resolutions as in Figure \ref{fig:f12}.
    }
  \label{fig:f13}
\end{figure*}

The fact that most of the returning mass flux is not associated with opposite
polarity magnetic flux indicates that it returns mostly by submerging
magnetic field rather than by flowing along downward directed fieldlines.
To quantify this aspect further we present in Figure \ref{fig:f13}a) the
inclination of flows and magnetic field relative to the horizontal
radial direction, i.e. the quantities $\gamma_v=\arctan(v_z/v_R)$ and
$\gamma_B=\arctan(B_z/B_R)$. We average these expressions in Evershed flow
channels separated into upflow and downflow regions, i.e. we consider
regions with $v_R>0$ and $v_z>250$ or $v_z<-250$ m$\,$s$^{-1}$. In upflow
regions both field and flow show a very similar inclination dropping
from about $30\deg$ at $R=8$~Mm to $10-15\deg$ at $R=15$~Mm.
For $R>15$~Mm the inclination for the magnetic field stays at around 
$10\deg$, while the flow inclination increases to $40\deg$. The situation
is different in downflow regions. Values for the flow inclination 
are $-40\deg$ in the inner and outer penumbra, while in the center 
of the penumbra values around $-20\deg$ are typical. The magnetic
field starts with an angle of $35\deg$ at $R=8$ and continues to point
upwards to about $R=14.5$~Mm. The minimum inclination the field reaches is
$-10\deg$ at $R=16$~Mm. Except for the cores of penumbral filaments
(equivalent to upflow regions) in the inner most $2/3$ of the penumbra,
the inclination angles of flow and field are quite different: Flows
in the penumbra are not just flows along the magnetic field, in particular not
in downflow regions. To illustrate this point more clearly we present in
Figure \ref{fig:f13}b) the average angle between flow and field given
by $\arccos(\vert \vec{v}\cdot\vec{B}\vert/(\vert\vec{v}\vert\,
\vert\vec{B}\vert))$. In the inner most $2/3$ of the penumbra the
angle between flow and field is $10-20\deg$ in upflow and
$25-60\deg$ in downflow regions. These values are larger than those evident
from panel a) since we measure here the average of the local misalignment, 
which is different from the misalignment of the average field and flow.

The inclination of flow and field agrees qualitatively with the
findings of \citet[][SOM]{Scharmer:etal:2011:sci}. They found that in locally
bright features (upflows) inclination angles of flow and field agree very well,
while the magnetic field was close to horizontal or weakly upward pointing
in downflow regions (locally dark features).

\citet{Franz:2011} concluded from Hinode data that at least $40\%$ of
penumbral downflows contain magnetic field with opposite polarity, 
similar to our finding. This agreement might be
coincidental as a substantial fraction of up- and down-flowing mass flux
still might be hidden in observations. \citet{Franz:2011}
speculated that this is a lower limit and possibly all downflows
have opposite polarity flux. 
We see only a moderate increase of this fraction with resolution 
(for the resolutions of $48\,[24]$, $32\,[16]$, $16\,[12]$~km we find 
$33\%$, $36\%$, $40\%$, respectively). In the
inner penumbra mass is returning beneath the photosphere by submerging
entire fieldlines (that still connect to the upper boundary) along the 
lateral downflow lanes of filaments without requiring opposite polarity flux 
to form \citep[see also Figure 19 in][]{Rempel:2011}. We see a moderate increase
in this number with overall extent of the penumbra for a fixed resolution
of $48\,[24]$~km from $29\%$ to $36\%$ (comparing the $\alpha=1.5$ and 
$\alpha=2.5$ cases). If
we extrapolate our results assuming that $100\%$ of downflows have opposite
polarity flux (for which we do not see any indication here), we would 
obtain values similar to those found by \citet{Sanchez-Almeida:2005:sunspot}.

From our models we expect that more than half of the returning mass
flux in the penumbra is found in regions that have the same polarity as the
umbra of the sunspot. Examples for such downflows were recently detected
by \citet{Katsukawa:Jurcak:2010}, although their connection to overturning
motions is not evident from their observations.

\begin{figure*}
  \centering 
  \resizebox{0.7\hsize}{!}{\includegraphics{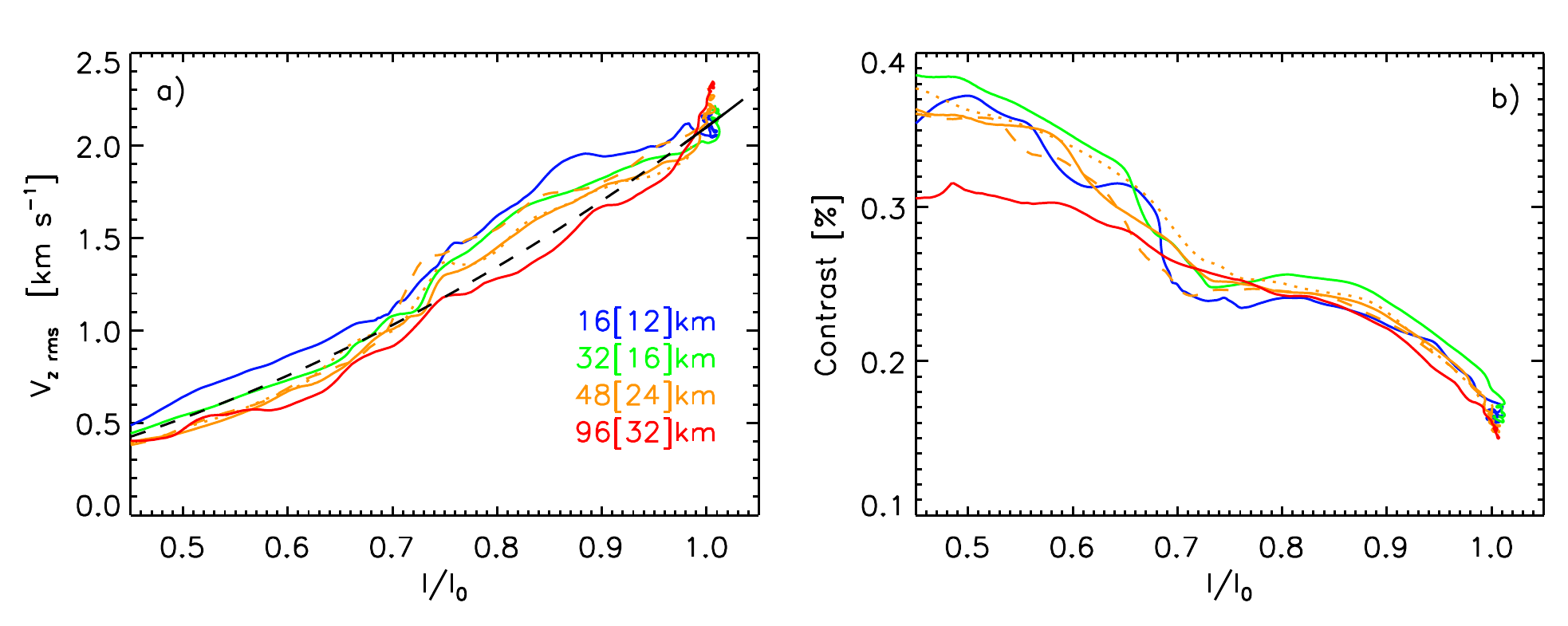}}
  \caption{Relation between azimuthally averaged bolometric intensity
    and vertical rms velocity (a) as well as rms intensity contrast (b)
    for different grid resolutions (color). In addition the dotted (dashed)
    lines show for comparison results with different top boundary
    conditions $\alpha=1.5\,(2.5)$ for the $48\,[24]$~km case. In panel a) the
    black dashed line indicates a relation of the form 
    $I\propto\sqrt{v_z^{rms}(\tau=1)}$.
  }
  \label{fig:f14}
\end{figure*}

\begin{figure*}
  \centering 
  \resizebox{0.7\hsize}{!}{\includegraphics{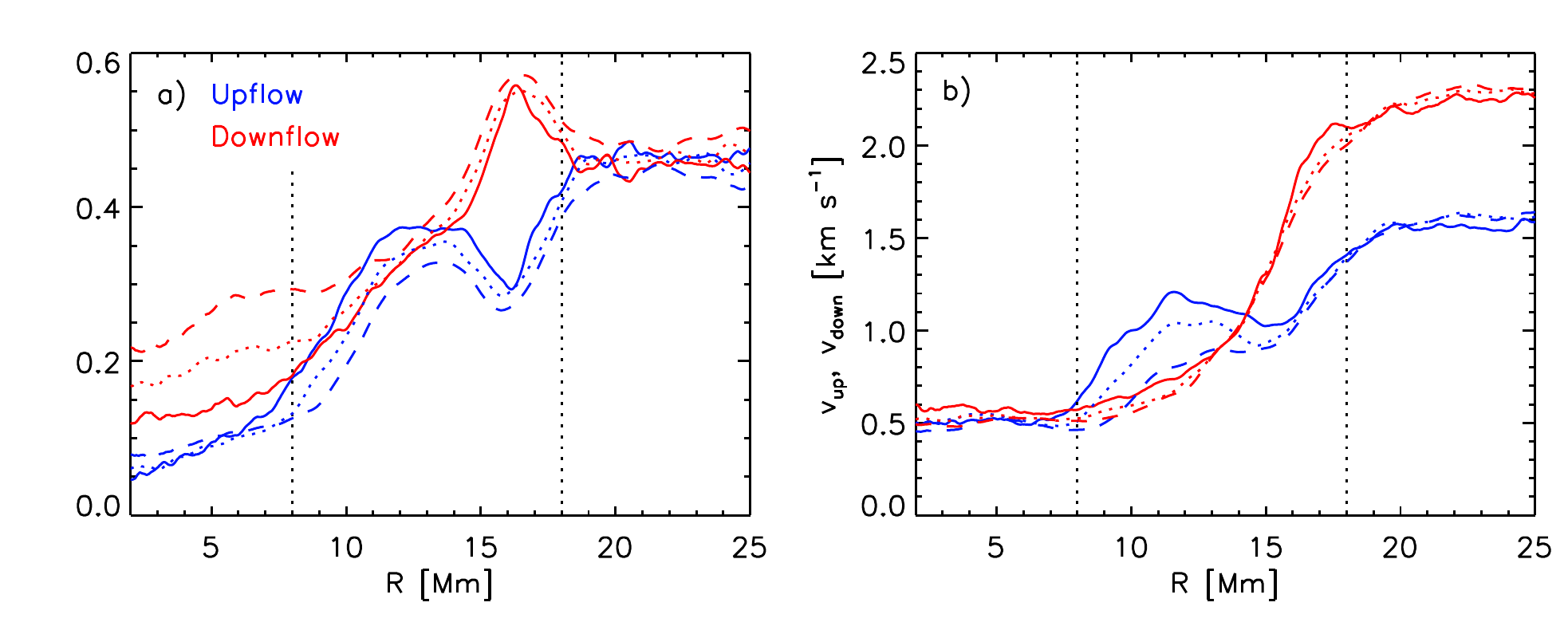}}
  \caption{a) Filling factors of regions with $v_z> 250$ m$\,$s$^{-1}$ (blue) 
    and regions with $v_z< -250$ m$\,$s$^{-1}$ (red). b) Average upflow (blue) 
    and downflow (red) velocity in those regions. All quantities are computed
    at the $\tau=1$ level. Solid, dotted, and dashed lines show results from 
    simulations with $16\,[12]$, $32\,[16]$, and $48\,[24]$~km resolution,
    respectively.
  }
  \label{fig:f15}
\end{figure*}

\subsection{Mass Flux associated with large scale flow component}
\label{sect:mflx_evershed}
The mass flux in the penumbra can be formally separated into a large scale 
mean flow with upflows in the inner and downflows in the outer penumbra and
laterally overturning motions. The former corresponds in the photosphere to
the Evershed flow. Note that this separation is only meaningful in terms of 
the overall mass flux balance. Individual flow elements have always a 
combination of radial outflow and lateral overturning motions.
To quantify both components of the mass flux we define 
\citep[see also][]{Rempel:2011} $m_z^+=\langle v_z\varrho>0\rangle$ and 
$m_z^-=\langle v_z\varrho<0\rangle$, where $\langle \ldots \rangle$ denotes
the azimuthal average. The unsigned vertical mass flux
is then given through $m_{\rm tot}=m_z^+-m_z^-$, while the locally unbalanced 
mass flux associated with radially separated regions of up and downflows (and
corresponding horizontal flows in between) is given by
$m_{\rm mean}=m_z^++m_z^-$. As a relative measure for the mass flux contained
in the azimuthal mean flow we consider 
\begin{equation}
  \eps=\frac{\int_{R_1}^{R_2}R\vert m_z^++m_z^-\vert \mbox{d}R}{ 
  \int_{R_1}^{R_2}R\vert m_z^+-m_z^-\vert \mbox{d}R}
\end{equation}
Evaluating this expression on a constant height surface about 350~km beneath
the $\tau=1$ level in the plage region (to make sure we stay below 
$\tau=1$ in the inner penumbra) we obtain with $R_1=6$ and 
$R_2=18$~Mm values of $15-16\%$ for grid resolutions from $48\,[24]$~km to 
$16\,[12]$~km. Similarly we obtain values within this range for the two 
simulations at $48\,[24]$~km resolution with different extent of the penumbra 
due to changes in the top boundary condition. Here we used a value of
$R_2=16$~Mm and $R_2=19$~Mm for the $\alpha=1.5$ and $\alpha=2.5$ cases to
consider the different extent of the region occupied by the Evershed flow
according to Figure \ref{fig:f2}. This is also in agreement with the value 
found previously by \citet{Rempel:2011} for the double sunspot simulation
of \citet{Rempel:etal:Science}. 

In order to evaluate
this quantity on a constant $\tau$ surface we have to replace $m_z$ with the 
mass flux normal to $\tau$ levels and include a geometric factor considering
projection effects due to inclined $\tau$ surfaces. Both effects are 
considered by using the expression  
$m_{\tau}=\vec{m}\cdot\nabla\bar{\tau}/\vert\partial_z\bar{\tau}\vert$
instead of $m_z$ at the respective $\tau$ level. Here
$\bar{\tau}$ denotes a horizontally smoothed $\tau$, since we did not
find a sufficiently balanced mass flux if we use $\tau$ at maximum resolution.
We compare here results that were obtained after convolving $\tau$ in the
horizontal direction with a Gaussian having a FWHM of $192$~km.
We find again comparable values for the resolution levels from $48\,[24]$~km 
to $16\,[12]$~km. Values of $\eps$ for the $\tau$ levels of $1.0\;(0.1)$ 
are $14-16\%\;(8-11\%)$ (note that this is relative to the unsigned total 
mass flux at the respective level, which is at the $\tau=0.1$ level about 
$25\%$ of that at $\tau=1$). The value at $\tau=1$ is very close to what 
we found on a constant height surface about $300$~km deeper, pointing toward
robustness of $\eps$ with regard to the height (or $\tau$ level) as well as
grid resolution.

\section{Overturning convection and visibility of convective signatures}
\label{sect:conv}
With the advent of magneto-convective models of sunspot penumbrae, the 
role of overturning convection in the penumbra has seen a 
controversial debate. 

On the theoretical side, overturning convection is the key process responsible 
for the energy transport and filamentation
\citep[see, e.g.,][]{Spruit:Scharmer:2006,Scharmer:Spruit:2006,Heinemann:etal:2007,Scharmer:etal:2008,Rempel:etal:Science,Rempel:etal:2009,Kitiashvili:etal:2009,Rempel:2011}. 

On the observational side, the evidence for overturning convection is 
controversial. While the existence of downflows and upflows is evident
\citep[see, e.g., the direct evidence by][]{DelToro:etal:2001,Franz:Schlichenmaier:2009}, their nature is debated: Do up- and downflows correspond to the 
radial endpoints of penumbral flow channels and are just the vertical 
component of the Evershed flow, or do individual penumbral
filaments show a flow pattern similar to that of granules with a substantial 
amount of mass flux turning over in lateral directions? With regard to the 
latter direct and indirect evidence was presented by \citet{Marquez:etal:2006,
Sanchez-Almeida:etal:2007,Ichimoto:etal:2007:sc,Rimmele:2008,
Zakharov:etal:2008,Bharti:etal:2010,Joshi:etal:2011,Scharmer:etal:2011:sci,
Scharmer:Henriques:2011}. Other investigations claim that these motions
do not exist \citep{Franz:Schlichenmaier:2009,BellotRubio:etal:2010}. 

\begin{figure*}
  \centering 
  \resizebox{0.7\hsize}{!}{\includegraphics{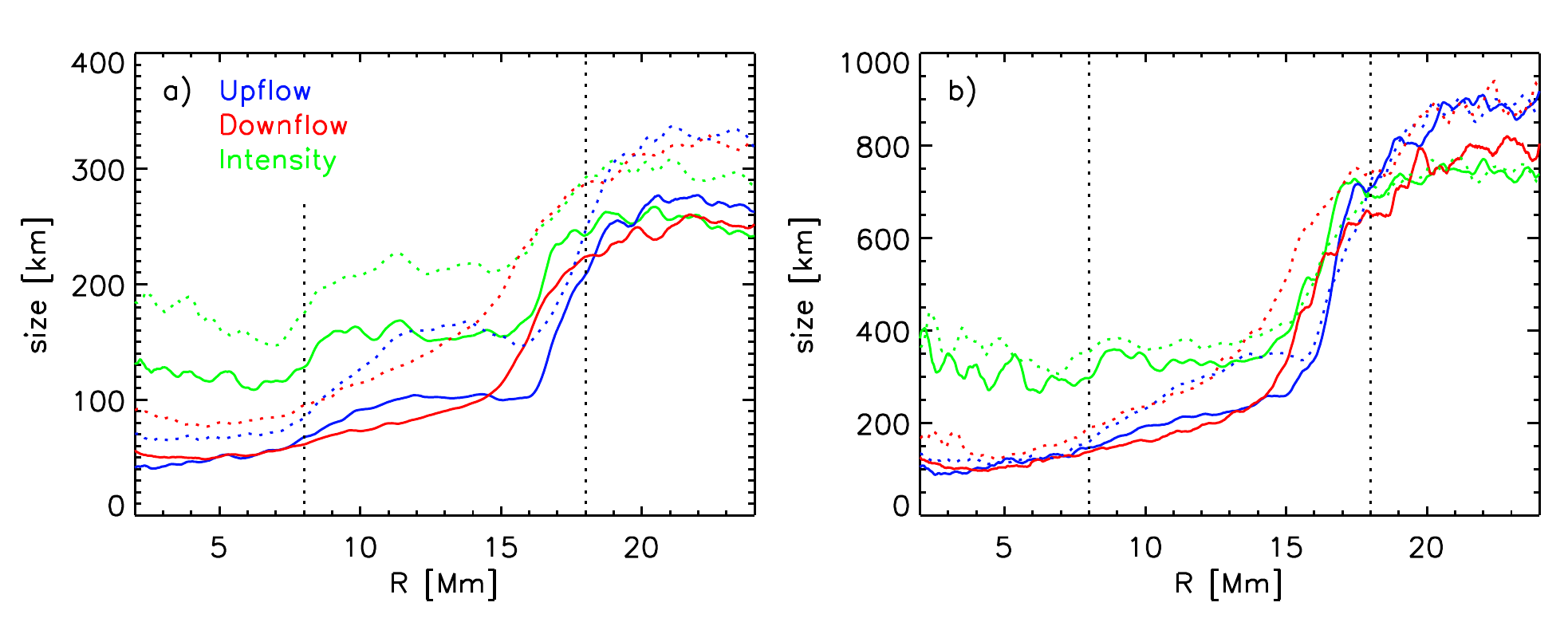}}
  \caption{Average azimuthal extent of intensity and velocity features: 
    regions with $v_z> 250$ m$\,$s$^{-1}$ (blue), regions with $v_z< -250$ 
    m$\,$s$^{-1}$ (red), and regions with intensity $0.05 I_{\odot}$ above 
    background (green). Panel a) shows the average, panel b) the area 
    weighted average. Solid (dotted) lines correspond to the simulations 
    with $16\,[12]$ ($32\,[16]$)~km resolution.  
  }
  \label{fig:f16}
\end{figure*}

\begin{figure*}
  \centering 
  \resizebox{0.7\hsize}{!}{\includegraphics{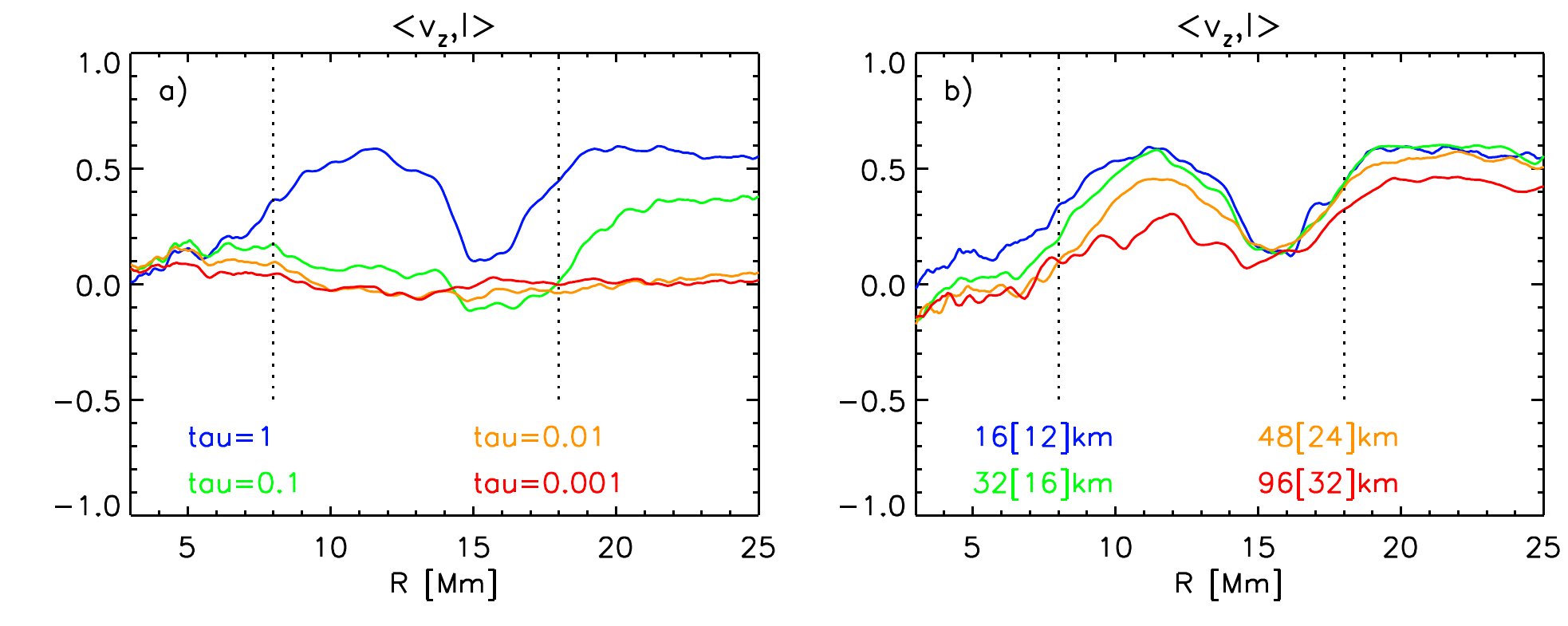}}
  \caption{Correlation between vertical velocity and bolometric intensity
    as function of a) height for a fixed resolution ($16\,[12]$~km) and b)
    resolution for a fixed height ($\tau=1$).}
  \label{fig:f17}
\end{figure*}

Figure \ref{fig:f14} quantifies the vertical rms velocity and the relative 
rms intensity contrast for different grid resolutions. In panel (a) the dashed 
line indicates an approximate relation of the form:
\begin{equation}
  I\propto\sqrt{v_z^{\rm rms}(\tau=1)}\,,
  \label{eq:e4}
\end{equation}
which was already found by \citet{Rempel:2011}. This relation means that 
about $50\%$ of the level of overturning convection found in the plage 
surrounding the sunspot is
required to maintain an intensity of about $0.7\,I_{\odot}$ typical for the
inner penumbra. This is possible since the rms contrast is with $25\%$ about
a factor of $1.5$ larger than the contrast in the plage ($17\%$). Note
that the contrast values are a few $\%$ higher in our gray simulations 
compared to a non-gray run.

We find again only a very moderate dependence on the grid resolution as well 
as on penumbra extent (dashed and dotted yellow line). Only the contrast of
the lowest resolution case shows a significant deviations in regions with
$I<0.7 I_{\odot}$. If there is any systematic variation at all, it is a very 
moderate increase of the rms velocity with 
increasing grid resolution. Overall the predictions about the amplitude of
overturning convection in the penumbra are very robust. Lower values 
for the amplitude of overturning motions would be only possible in combination
with a substantially higher contrast to satisfy energy flux constraints.

While the overall amount of overturning convection as characterized by the 
vertical rms velocity is robust, there are differences in how much up- and
downflows contribute. To characterize their contribution further we show in
Figure \ref{fig:f15}a) the filling factors of regions with more than 
$250$~m$\,$s$^{-1}$ upflow (blue) and less than $-250$~m$\,$s$^{-1}$ downflow 
velocity (red). Filling factors of up- and downflows are comparable in the 
plage region outside the sunspot with values of about $45\%$. The outer 
penumbra shows a clear preference for downflows with filling factor of $55\%$, 
while upflows occupy only $30\%$ of the area. In the inner penumbra the 
upflow filling factor exceeds the downflow filling factor in the highest 
resolution case and is comparable for the other cases. In Figure 
\ref{fig:f15}b) we show the respective average velocities in up- and downflow 
regions. In the plage region 
downflows dominate over upflows by a factor of about $1.5$. This is mostly an
optical depth effect since we show quantities on the $\tau=1$ surface. Since 
up- and downflow velocity amplitude decline with height, the elevated
$\tau$ levels in upflow regions sample lower velocities than the depressed
$\tau$ levels in downflow regions, while the difference on a constant height
surface is typically less pronounced. This ``apparent redshift'' 
is not to be confused with the well known convective blueshift that originates 
in unresolved observations from the different intensity weighting of up- and 
downflow regions dependent 
on the the temperature sensitivity of the considered spectral line. In the
inner penumbra upflow velocities dominate over downflows by
a factor of up to $1.7$ (despite the fact that also here the elevated $\tau$ 
levels in upflow regions produce a bias toward downflows). For the currently
explored resolution range we do not yet see a convergence of filling factors
as well as mean up- and downflow velocities. It is likely that the trend toward 
upflow dominance in the inner penumbra will continue somewhat further with 
increasing resolution.

\begin{figure*}
  \centering 
  \resizebox{0.95\hsize}{!}{\includegraphics{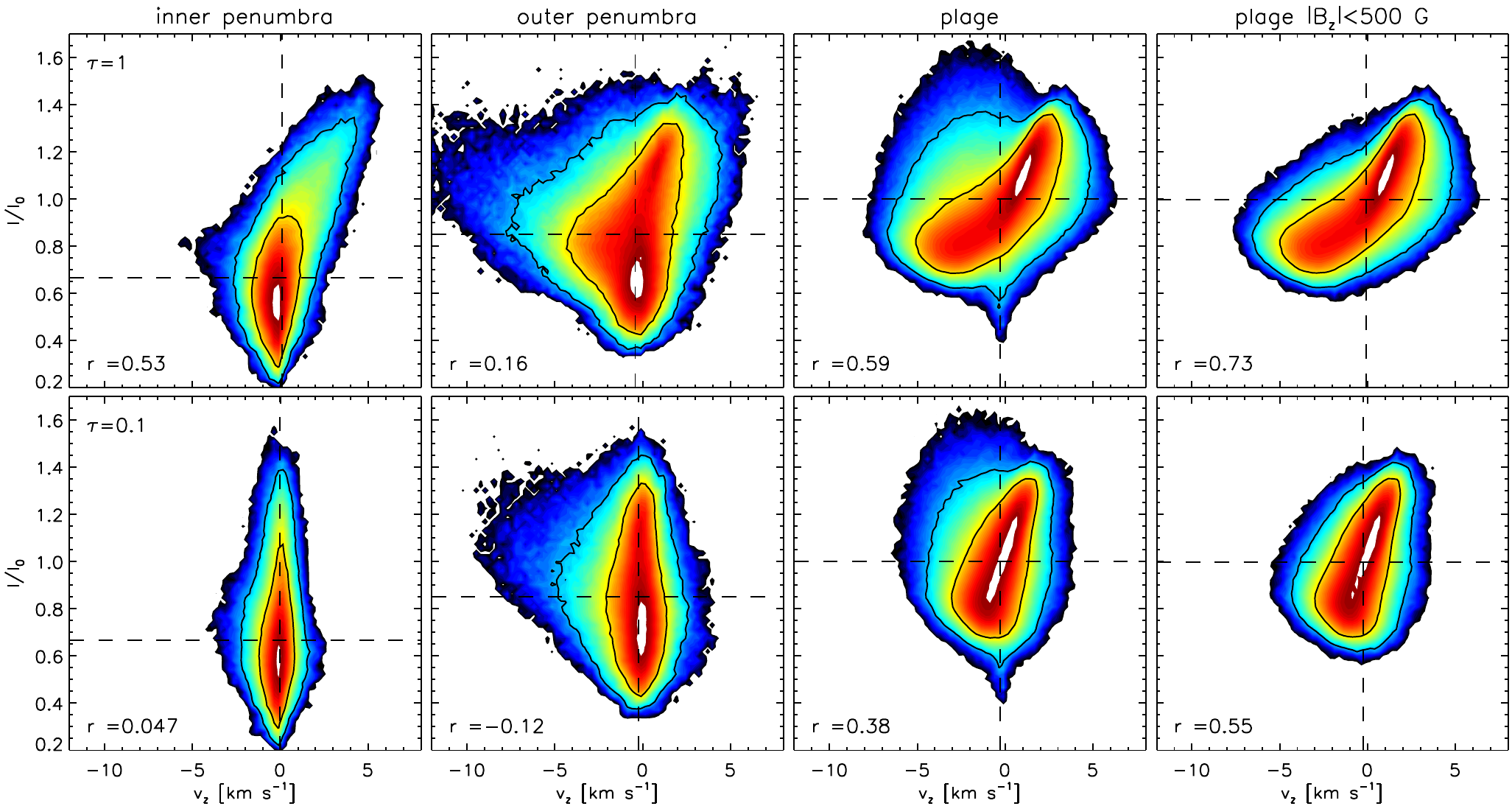}}
  \caption{Bivariate probability density functions for intensity and vertical 
    velocity computed for the simulation with $16\,[12]$~km 
    resolution. We defined four regions: The inner penumbra is from $R=10$ to 
    $R=14$~Mm, the outer penumbra from $R=14$ to $R=18$~Mm, and plage is
    everything outside $R=20$~Mm. In addition we computed the distribution
    function for plage regions with less than 500 G vertical field. The top
    row is computed with the vertical velocity at $\tau=1$, the bottom row
    with the velocity at $\tau=0.1$. We show the PDFs on a logarithmic scale 
    in the range from $0.001$ to $1$ of the maximum value, contour lines show 
    the levels of $0.1$ and $0.01$. Dashed horizontal and vertical lines 
    show the average values for the considered sub regions.
  }
  \label{fig:f18}
\end{figure*}

In Figure \ref{fig:f16} we present the widths of intensity, up- and downflow 
features as function of radius. To this end we determine the average azimuthal
extent of regions with intensity larger than $0.05 I_{\odot}$ compared to the
background that is defined through a smoothed intensity profile (Gaussian 
with a FWHM of $1.5$~Mm). Up- and downflows are again defined through regions
 with more than $250$~m$\,$s$^{-1}$ velocity amplitude. We consider here the 
following measures for the widths
\begin{equation}
  \langle s(r)\rangle_n=\frac{\sum_i s_i(r)^{n+1}}{\sum_i s_i(r)^n}\;,
\end{equation}  
where $s_i(r)$ are the detected widths at a given radial position. For
$n=0$ we obtain the regular average, while $n=2$ can be considered as an
``area weighted average'' if we make the assumption that the area of a feature 
scales with $s_i^2$. Figure \ref{fig:f16} shows in panel a) 
$\langle s(r)\rangle_0$ and in panel b) $\langle s(r)\rangle_2$. Regardless
of the adopted measure we find that the widths of intensity and
flow features are comparable in the plage region, while flow features
are systematically smaller in the penumbra by a factor of up to 2. We do
not see convergence of $\langle s(r)\rangle_0$ with grid resolution (we show 
here the $16\,[12]$ and $32\,[16]$~km cases by solid and dotted lines) in
neither penumbra nor plage due to small scale features that dominate by 
numbers in both regions. The area weighted average $\langle s(r)\rangle_2$ 
shows better convergence in particular for the width of intensity features. 
The most dramatic changes occur in flow features in the mid to outer penumbra.
The latter indicates that even higher resolution than 16~km might be required
to capture the width of flow features in the penumbra properly and that the 
current width of about 100 (200)~km has to be considered as an upper limit
for the average (area weighted average) width.  

In Figure \ref{fig:f17} we present the $v_z-I$ correlation, which is very
often used to characterize convective energy transport.
The $v_z-I$ correlation reaches in the inner penumbra values similar to
those found outside the sunspot (peak at around 0.6). Interestingly, the
correlation vanishes in the outer penumbra. We further find that a 
significant correlation is only present with the velocity at $\tau=1$
inside the penumbra, whereas $\tau=0.1$ leads also to a very moderate 
($0.3-0.4$) outside the sunspot. 
We find here again only little resolution dependence in the range from 
$48\,[24]$ to $16\,[12]$~km. A $v_z-I$ correlation in the inner penumbra
comparable to the plage surrounding the sunspot and a very weak correlation
in the outer penumbra are captured well starting from a resolution of
$48\,[24]$~km. 

Note that the drop of the correlation in the outer penumbra is due to
photospheric effects. If we compute instead the correlation between $m_z$
and $(e_{\rm int}+p)/\varrho+1/2 v^2$ (here $e_{\rm int}$ and $p$ denote the 
internal energy and gas pressure) on a constant height surface located 
about $350$~km
beneath average $\tau=1$ in the plage region, we find a constant value of 
$0.65$ for $R>8$~Mm. This indicates that rather moderate values of the $I-v_z$ 
correlation in the photosphere cannot be taken as an argument against 
convective energy transport, in particular not if strong magnetic field is 
present. 

\begin{figure*}
  \centering 
  \resizebox{0.95\hsize}{!}{\includegraphics{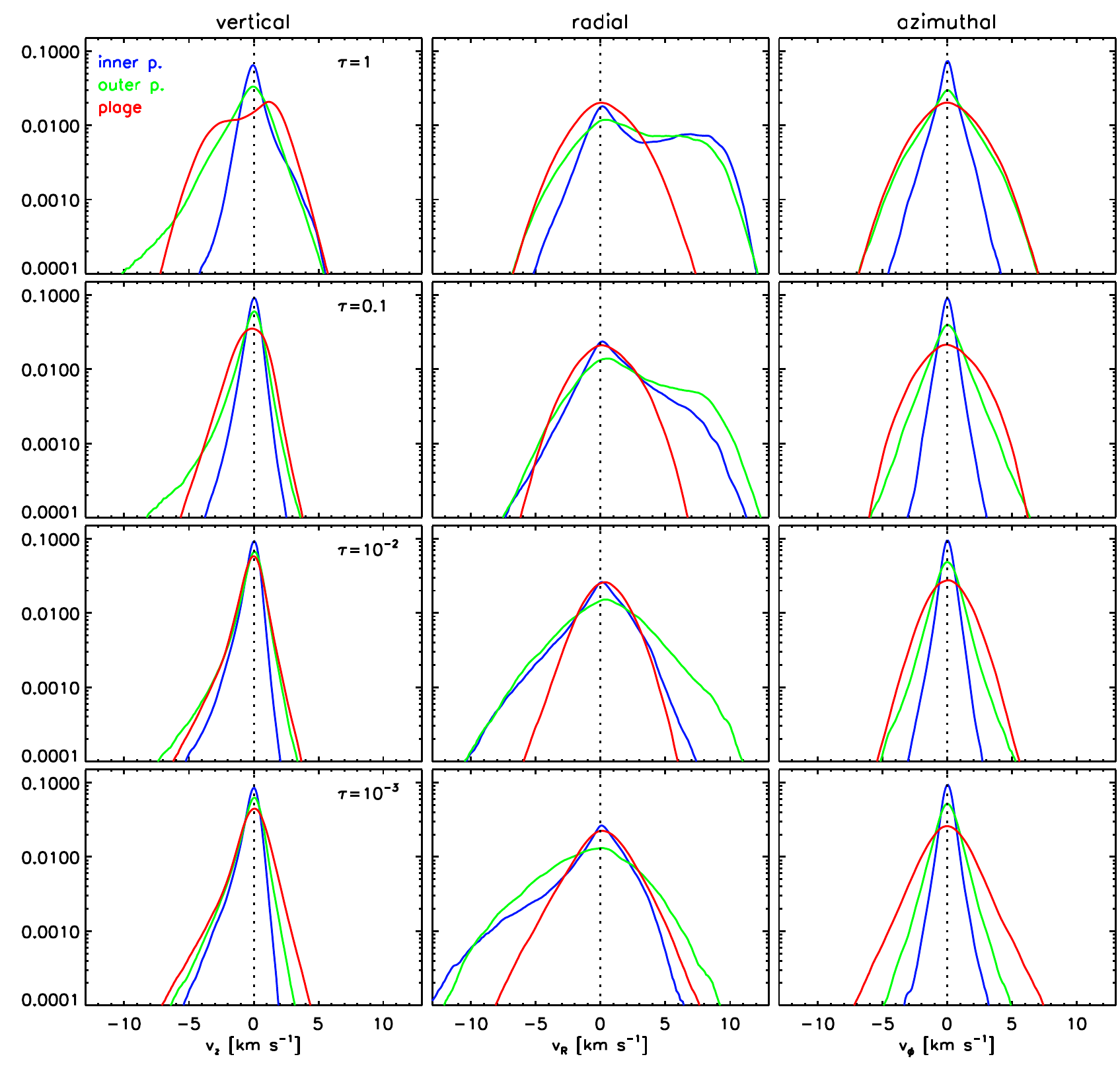}}
  \caption{Velocity distribution functions in inner penumbra (from $R=10$ to 
    $R=14$ Mm, blue), outer penumbra (from $R=14$ to $R=18$ Mm, green) and 
    plage ($R>20$ Mm, red). Left to right we show the distribution functions
    for vertical, radial, and azimuthal flows; top to bottom the levels of
    $\tau=1$, $0.1$, $10^{-2}$, and $10^{-3}$.}
  \label{fig:f19}
\end{figure*}
 
\section{Probability density functions of photospheric velocity field}
\label{sect:statistics}
In Section \ref{sect:conv} we discussed the correlation between intensity
and vertical velocity. We found comparable values ($\sim 0.6$) in the inner 
penumbra and the strong plage surrounding the sunspot, while the correlation 
disappears in the outer penumbra. In Figure \ref{fig:f18} we show for
the highest resolution case ($16\,[12]$~km) bivariate probability density 
functions for vertical velocity and intensity,
computed (left to right) for the inner penumbra ($R=10$ to $R=14$~Mm), the 
outer penumbra ($R=14$ to $R=18$~Mm), plage ($R>20$~Mm) as well as plage 
regions with $\vert B_z\vert < 500$~G. We use here wider masks
for the inner and outer penumbra than in Section \ref{sect:Bevershed} since
we want to include all of the fast downflows at the outer edge of the
penumbra. The PDFs are computed for absolute intensity values (rather
than fluctuations about the background) to allow for a direct comparison of
intensity values between penumbra and plage.

Comparing first the plage region and plage with  $\vert B_z\vert < 500$~G, we 
see that the  moderate correlation of $0.59$ is mostly due to a 
population of very bright downflows that are associated with strong field.
Masking out these regions increases the correlation to $0.73$. The correlation
for the inner penumbra has a value of $0.53$ very close to that of plage,
however, the shape of the PDF is substantially different. While the
correlation in the plage regions results in about equal parts from
bright upflows and dark downflows, it is in the inner penumbra mostly
due to bright upflows, some with brightness values reaching $1.4 I_{\odot}$. 
Similarly to the plage region, the decorrelation in the 
outer penumbra is due to a strong population of bright downflows. In 
particular, most supersonic downflows have a brightness comparable to 
$I_{\odot}$. Most of these features have also strong opposite polarity
magnetic field, indicating that the mechanism for the brightness enhancement
is likely similar in penumbra and plage.

The less pronounced tendency for downflows to be darker in the inner penumbra 
is in part due to the fact that the typical width of flow features is about
half the width of intensity features according to Figure \ref{fig:f16}.
While the upflows are mostly located centrally in bright patches, downflows
are located near the edge and are partially found in bright and dark features.

Computing the PDFs based on the vertical velocity at $\tau=0.1$ does not
lead to any significant correlation in the penumbra, while the plage
region shows values of $0.38$ and $0.55$ excluding strong field regions.
The core of the PDFs leads in most cases to a stronger
correlation. This implies that smoothing the data increases 
the overall level of correlation found for both penumbra and plage. For example
convolving the data by a Gaussian with FWHM of $150$~km leads for $\tau=1$
to correlations of $0.63$, $0.19$, $0.72$, and $0.82$. These values are
comparable to those recently found by \citet{Scharmer:Henriques:2011} for the
interior penumbra and quiet sun. We see overall a good agreement 
with their results, although differences in the shape of the $I-v_z$
probability density functions for penumbra and plage between their observations
and our simulations exist. They did find quite comparable results for the
CI 5380 and FeI 6301 lines, while we see a rather strong height dependence
for the probability density functions and resulting correlations between
$\tau=1$ and $\tau=0.1$. 
The bivariate probability density functions do not show substantial 
differences between the  $48\,[24]$~km and  $16\,[12]$~km cases in terms of 
the features discussed above.

PDFs for $v_z$, $v_R$ and $v_{\Phi}$ are presented in Figure \ref{fig:f19}.
At $\tau=1$ the PDF for $v_z$ peaks in the plage region near 
$v_z=1.5$~km$\,$s$^{-1}$ (the ``typical'' upflow velocity), while the PDFs for 
inner and outer penumbra peak 
near $v_z=0$~km$\,$s$^{-1}$. The PDF in the inner penumbra is skewed toward 
upflows, the PDF in the outer penumbra has a far extending tail of 
fast supersonic downflows. Toward lower $\tau$ values the shape of
the PDF for all three regions becomes similar, while the PDF in inner and
outer penumbra remains more narrow compared to the plage region.
Note that the fast downflow wing in the outer penumbra extends toward
higher layers than the fast upflow wing in the inner penumbra, which is
only present at $\tau=1$. A similar behaviour was seen by 
\citet{Scharmer:Henriques:2011} who found that downflows have a similar 
strength in CI 5380 and  FeI 6301, while upflows are stronger in the deeper 
forming CI 5380 line.

The PDF for $v_R$ is broader in the penumbra compared to plage at all
$\tau$ levels shown. For $\tau=1$ and $0.1$ it is skewed toward outflows,
while the inflow part matches up with that of the plage region; toward
$\tau=0.001$ it is skewed toward inflows, while the outflow
part matches up that of the plage region. In the inner penumbra the PDF
shows a secondary peak at $\tau=1$ at an outflow velocity of about 
$7-8$~km$\,$s$^{-1}$, corresponding to the ``typical'' Evershed flow velocity
within flow channels (the azimuthal average is about a factor of two smaller).

At $\tau=1$ the PDF for laterally overturning motions in outer penumbra
and plage are similar, while the inner penumbra shows a more narrow PDF.
In higher layers the PDFs for penumbra are more narrow than the
PDF for plage. 

Observed distribution functions for the vertical velocity in the range
from $-2$ to $2$ km$\,$s$^{-1}$ were shown by 
\citet{Franz:Schlichenmaier:2009}. 
They found in the quiet sun that upflows dominate over downflows for
all velocity intervals. In the penumbra (they did not distinguish between
inner and outer penumbra) upflows dominate for 
$\vert v_z\vert<400$~m$\,$s$^{-1}$, downflows for faster velocities. The
overall width of the PDF in the penumbra was about half of the value found
in the quiet sun. While our results certainly show about a factor of 2
difference in the widths (see also the vertical rms velocity in
Figure \ref{fig:f14}), other details might depend also on the
utilized spectral line as well as resolution and require likely a 
comparison through forward modeling.   

\begin{figure*}
  \centering 
  \resizebox{0.95\hsize}{!}{\includegraphics{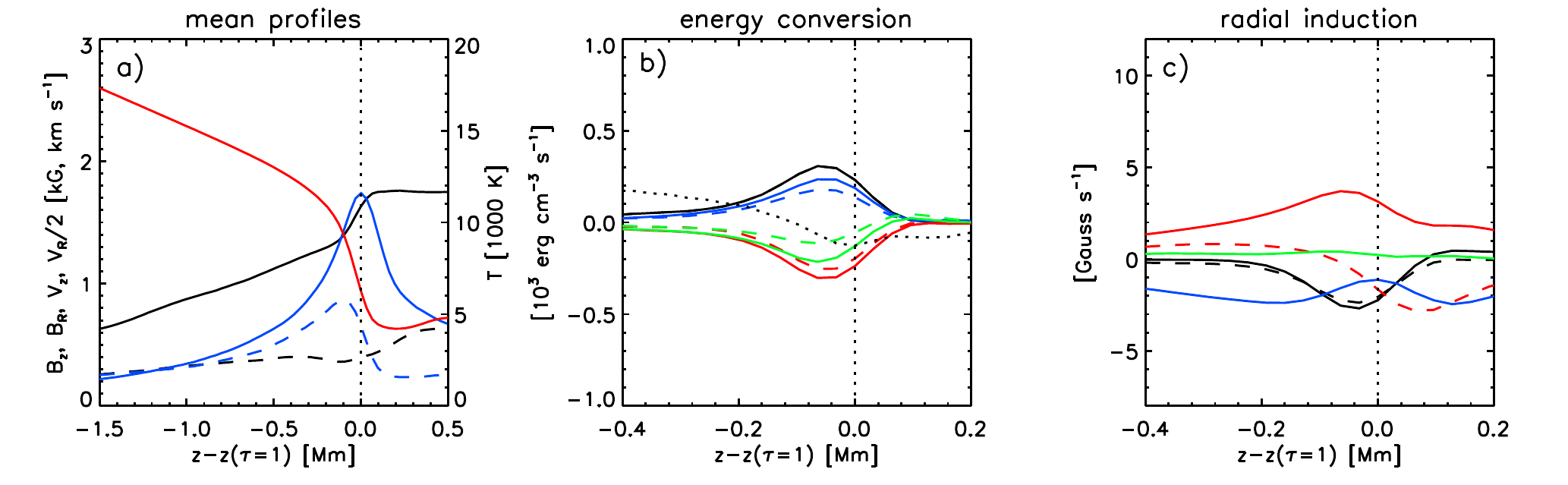}}
  \resizebox{0.95\hsize}{!}{\includegraphics{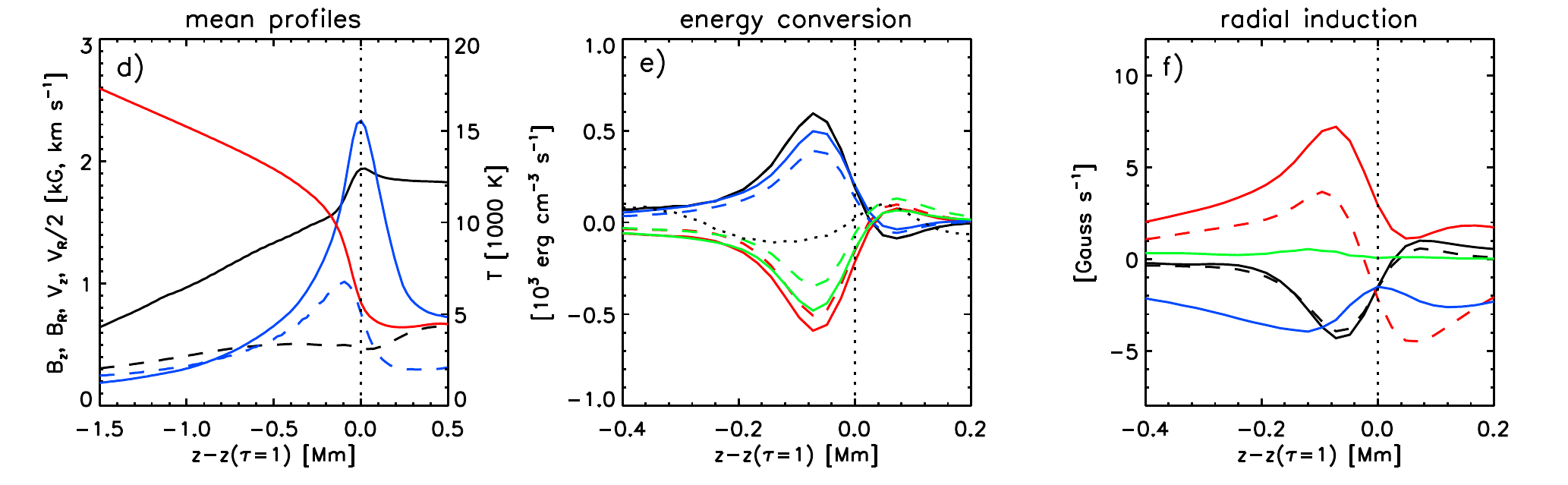}}
  \resizebox{0.95\hsize}{!}{\includegraphics{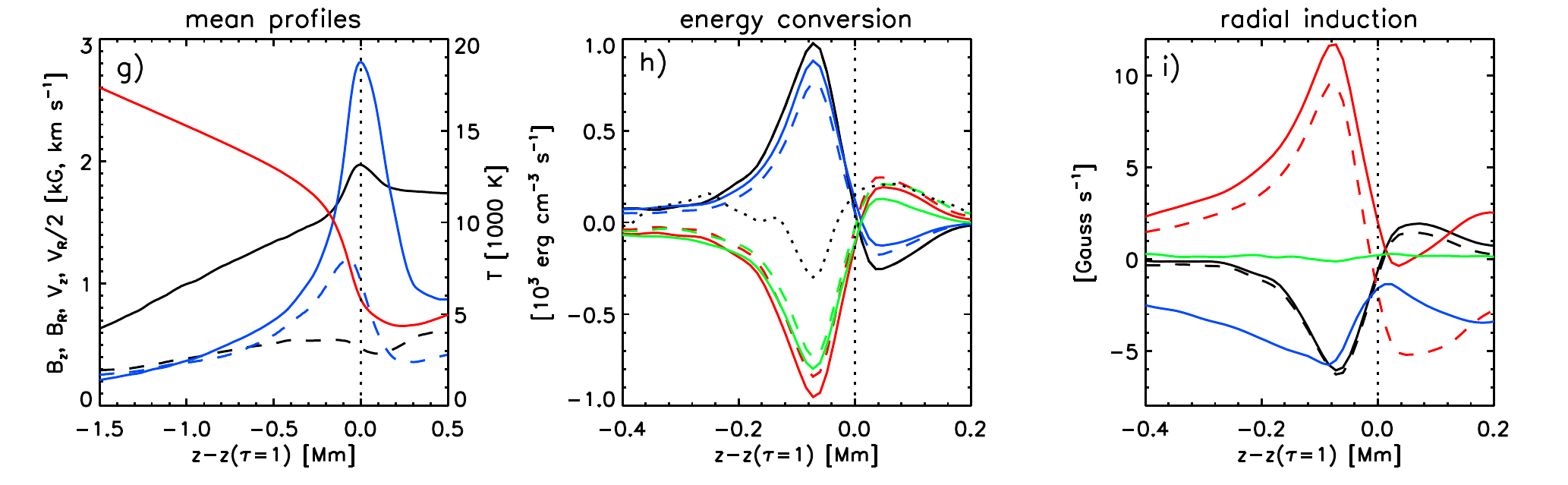}}
  \caption{Horizontal averages between R=11 and R=14.5 Mm over regions with
    $v_R >0$ and $v_z>0$ as function of height following \citet{Rempel:2011}. 
    The left panels (a,d,g) show
    $B_R$ (black, solid), $B_z$ (black, dashed), $1/2v_R$ (blue, solid), 
    $v_z$ (blue, dashed), and T (red). The center panels (b,e,h) show the
    energy conversion rates by different force components: pressure/buoyancy
    forces in vertical direction (black), Lorentz force in vertical direction
    (red), Lorentz force in horizontal direction (blue), and acceleration force
    in horizontal direction (green). Dashed lines refer to simplified 
    expressions as explained in the text. The right panels (c,f,i) show
    different contributions in the radial induction equation: stretching
    (red), advection (black), divergence (blue), resistivity (green). The
    red dashed line shows the contribution from the vertical subsurface 
    velocity shear $\langle B_z\partial_z v_R\rangle$. The numerical grid
    resolution is increasing top to bottom: panels (a,b,c) $96\,[32]$~km,
    panels (d,e,f) $48\,[24]$~km, and panels (g,h,i) $16\,[12]$~km.
  }
  \label{fig:f20}
\end{figure*}
  
\section{Resolution dependence of subsurface energy exchange}
\label{sect:subsurf}
In Figure \ref{fig:f20} we analyze subphotospheric processes responsible for the
Evershed flow as well as the horizontal magnetic field found in filament
flow channels. Similarly to \citet{Rempel:2011} we focus here on regions with
$v_z>0$ and $v_R>0$ in the inner penumbra, where the strongest driving of
the Evershed flow takes place. 

The left panels (a,d,g) present averages of magnetic field,
velocity and temperature. Apart from the dramatic change of the Evershed flow
speed we also find that the enhancement of $B_R$ at $\tau=1$ is less pronounced
at lower resolution, consistent with Figure \ref{fig:f8}. 

The middle panels
(b,e,h) show energy conversion terms that play a central role in the maintenance
of the Evershed flow. As discussed in detail in \citet{Rempel:2011} pressure
buoyancy driving takes place in upflow regions, while downflows are close
to hydrostatic, which is opposite to the situation in non-magnetic convection.
The pressure driving in upflows is diverted by the magnetic field into the
horizontal direction where the acceleration takes place, while the pressure 
force itself does not have a strong component in the horizontal direction.
To quantify the influence from grid resolution on this process we show the
work by the vertical component of the pressure/buoyancy force (black), work by
the vertical component of the Lorentz force (red), work by the radial 
component of the Lorentz force (blue) and radial component of the acceleration
force (green). Dashed lines correspond to simplified expressions that were
discussed
further in \citet{Rempel:2011}. The dotted line indicates the work by the
radial component of the pressure force multiplied a factor of 10 for better
visibility. Regardless of resolution we find the same balance, only the overall
amplitudes of the various terms increase with increasing resolution. A quadratic
increase with the peak Evershed flow velocity is expected, since the 
acceleration term is given in leading order by 
$-\langle\varrho v_R v_z\partial_zv_R\rangle$ and $v_z$ itself does not change
with resolution (tied to the fixed amount of energy transported). 

The panels on the right (c,f,i) show the contributions from the stretching 
term (red) advection term (black) and divergence term (blue) in the
induction equation for the radial field component. Here it is most striking
that the stretching term and in particular the contributions from
$B_z\partial_zv_R$ (red, dashed) triples in amplitude over the resolution
range explored. As discussed in detail in \citet{Rempel:2011}, there is a
strong connection between the Evershed flow and the horizontal magnetic field 
in a thin boundary layer beneath the $\tau=1$ surface. The subsurface shear 
profile of the Evershed flow induces strong horizontal field, which is 
distributed along the $\tau=1$ level by overturning convection and leads to 
a thin ``outer shell'' of the filament with strongly enhanced horizontal field.
This boundary layer is also the place where strong magnetic
tension forces are in balance with vertical pressure and horizontal 
acceleration forces (see also Figure \ref{fig:f21}). 
This strong linkage between $B_R$ and $v_R$ is also evident 
from Figures \ref{fig:f4} and \ref{fig:f8}, in which 
both the Evershed flow and horizontal magnetic field show the strongest 
resolution dependence with a trend toward faster flows and stronger field
at higher resolution.  

\section{Discussion}
\label{sect:discussion}
\subsection{Role of top boundary condition}
We found a very strong influence of the magnetic top boundary condition
on the extent of the penumbra. We do not obtain
a penumbra at all using a potential field extrapolation since the horizontal
periodicity implies an asymptotically vertical field. This is in agreement with
earlier results by \citet{Rempel:etal:Science}, who only found extended
penumbrae in between nearby opposite polarity spots. 
This rather puzzling result requires further investigation, we see three 
possible implications: 

1. This result is entirely due to the horizontal periodicity that
is chosen for computational convenience. 

2. Other physical processes are still 
missing or a insufficiently resolved, like turbulent pumping as suggested by 
\citet{Montesinos:Thomas:1997,Brummell:etal:2008}.
While processes along the lines of turbulent pumping are certainly present
in our simulations \citep[see, also][]{Rempel:2011} and can be associated
with the opposite polarity flux present in the outer penumbra, we did not find
any indication of a strong resolution dependence here that would indicate that 
these processes could be underestimated. 

3. Hysteresis from our initial state could prevent the formation of a penumbra.
To test this we conducted additional experiments with a sunspot 
model similar to our $48\,[24]$~km resolution case, but in a larger
$73.728\times 73.728\times 9.216$~Mm$^3$ domain (this model has been described
also in \citet{Rempel:2011:moat}). After running that simulation for a total
of 24 hours of solar time with our $\alpha=2$ boundary condition, we switched 
back to $\alpha=1$ (potential field). Within about 30 minutes the entire 
penumbra including the associated Evershed flow disappeared. After running
for 2 hours with the potential field boundary condition we switched back to the
$\alpha=2$ case and the entire penumbra inclusive Evershed flow reappeared in
about 30 minutes with properties essentially identically to the previous.
This indicates that there is no hysteresis from within the convection zone
in our numerical setup. This is not necessarily in contradiction to observed
hysteresis effects in sunspots if the hysteresis would be caused by the global
overlying magnetic field structure. Independent evidence against hysteresis
from our initial state comes from flux emergence simulations
\citep{Cheung:etal:2010}, which also fail in producing a penumbra if a potential
field boundary is used in a periodic domain. Overall this
indicates that the coronal magnetic field overlying sunspots has a potential
feedback on penumbral structure. This has been proposed by 
\citet{Liu:etal:2005,Deng:etal:2005}, who observed changes in penumbral
structure associated with solar flares. Our simulations certainly indicate
that such feedback is possible and that in response parts of penumbrae could
disappear or form on a time scale of about 30 minutes. Additional support
for a key role of the magnetic field overlying the sunspot comes from
\citet{Shimizu:etal:2012}, who discovered in a developing active region a 
chromospheric precursor of penumbra formation. They observed the formation 
of a magnetic canopy in the chromosphere before the penumbra formation took 
place in the photosphere. The penumbra reached in the end an extent similar to 
the previously observed chromospheric signature. Their observation indicates 
that the magnetic field at chromospheric levels likely plays an active role 
in the penumbra formation process.  
   
\subsection{Convergence with regard to numerical resolution}
Overall we find that penumbral fine structure is robust with regard to grid
resolution over the explored range from $96\,[32]$ to $16\,[12]$~km. Even 
our lowest resolution case (which could be considered as an intentional 
experiment of under-resolving penumbral structure) gives still a good 
qualitative agreement. Quantitative agreement requires better than
$48\,[24]$~km resolution. This agreement is in terms of the structure and
correlations of quantities in the penumbra in particular Evershed flow 
channels, opposite polarity magnetic flux and its relation to downflows,
the level of overturning convection and the subsurface magneto convective
processes responsible for the filamentation and driving of the Evershed flow.
All resolution levels considered are not sufficient to resolve turbulence 
within flow channels. However, we also find strong magnetization of Evershed
flow channels as a robust result. Therefore it is doubtful that a substantial
amount of turbulence would develop at photospheric levels. Studies of sunspot 
penumbrae at higher resolution than considered here would be only meaningful 
if a nested or adaptive grid is used that allows for high local resolution, 
while the domain remains large enough to capture an entire sunspot.

We do not use any explicit viscosity or magnetic resistivity in our 
simulations. The unavoidable hyper-diffusivity used is resolution and scale
dependent. It scales at least linear with grid spacing near discontinuities 
or regions with monotonicity changes, but has higher order or is even switched 
off in well resolved regions. Our experiments on numerical resolution
implicitly explore the dependence of the solutions on hyper-diffusivities. 
From power spectra we estimate that
hyper-diffusion affects mostly features that are resolved by less than 6
grid points or have a scale of less than $100$ km in our simulation with 
$16\,[12]$~km resolution. This is consistent with our findings that details
of up- and downflow structures that exist on these scales in the penumbra are 
not fully converged yet.

\subsection{Comment on diffusion}
\label{sect:comment_diff}
We found in our simulations a rather bright umbra of $0.3\,I_{\odot}$ for a 
sunspot with $3.3$~kG central field strength (see Figure \ref{fig:f4}). 
Further investigation shows that the umbral brightness is influenced to some 
degree by numerical diffusivities, in particular mass-diffusion. Using a 
modified numerical approach with a lower mass-diffusion rate reduces the 
umbral brightness by $0.05-0.1\,I_{\odot}$. This difference is mostly due to
a different number density of umbral dots as details of diffusivities influence
the magnetic field threshold at which umbral dots become completely suppressed. 
We did not find a significant influence on other aspects of the solution. 
Numerical mass-diffusion mimics to some degree the effects of ambipolar 
diffusion, which indicates that some detail of umbral dots could be 
influenced by multi-fluid effects.  

\subsection{Visibility of convective signatures in real observations}
The vertical rms velocity depends very weakly on grid resolution, showing 
a very moderate increase with increasing grid resolution. The average downflow 
velocity is larger than the upflow velocity in plage and the outer penumbra 
(by about a factor of $1.5$), but in the inner penumbra upflows dominate over 
downflows by about a factor of $1.7$. In the outer penumbra downflows have a 
substantially larger filling factor ($55\%$) than upflows ($30\%$), while
the upflow filling factor dominates in the inner penumbra by a small margin.
These differences between up and downflow regions become more pronounced with 
higher grid resolution in the inner penumbra. The consequence of these 
asymmetries is that upflows (downflows) tend to be easier to observe in 
the inner (outer) penumbra, i.e. with insufficient observational
resolution one should expect to see a pattern of upflows in the inner and
downflows in the outer penumbra. That raises the question of what a sufficient
observational resolution is in this context. \citet{Bharti:etal:2011:conv} used
a non-gray version of the model in $32\,[16]$~km grid resolution to address this
question through forward modeling. They focused mostly on a few filaments 
in the inner penumbra and found that the upflow would be clearly visible
at $0.14"$ observational resolution, while the visibility of the lateral 
downflows depends 
strongly on the spectral lines used. They found only $200$~m$\,$s$^{-1}$ in the 
Fe 7090 line, while the CI 5380 line yielded up to $800$~m$\,$s$^{-1}$ downflow
velocity. Based on the results presented here these numbers are likely still on
the optimistic side, since the azimuthal extent of flow structures is still
decreasing with increasing grid resolution (see Figure \ref{fig:f16}) and
\citet{Bharti:etal:2011:conv} did not consider additional effects from stray 
light and potential line blends that complicate the situation in real 
observations. In that sense we do not see at this point a direct conflict
with non-detection of overturning motions in penumbra filaments such as
\citep{Franz:Schlichenmaier:2009,BellotRubio:etal:2010}. Furthermore,
recent observations in the CI 5380 line by 
\citet{Joshi:etal:2011,Scharmer:etal:2011:sci} found direct evidence for
such motions, although their analysis depends strongly on stray light
removal and complications due to the projection of the Evershed flow
on the line of sight.  

Overall this indicates that observations are approaching the required 
resolution, which makes the outlook on the next generation solar telescopes
such as {\em NST}, {\em Gregor} and {\em ATST} very promising. A recent
example of new details that can be learned from comparing high resolution
observations with state of the art numerical models was presented by 
\citet{Steiner:etal:2010} using {\em Sunrise} observations of the quiet sun.
On the side
of simulations the presence of overturning motions is robust as they are
linked to very fundamental energy flux constraints. However, details still
change, indicating that also here higher grid resolution will be required for 
an in depth comparison with observations through forward modeling. For the 
current simulations spectral line features that result from structures on 
scales of $100$~km or less should be interpreted with caution, since details 
are likely influenced by numerical diffusion.  

\begin{figure*}
  \centering 
  \resizebox{0.75\hsize}{!}{\includegraphics{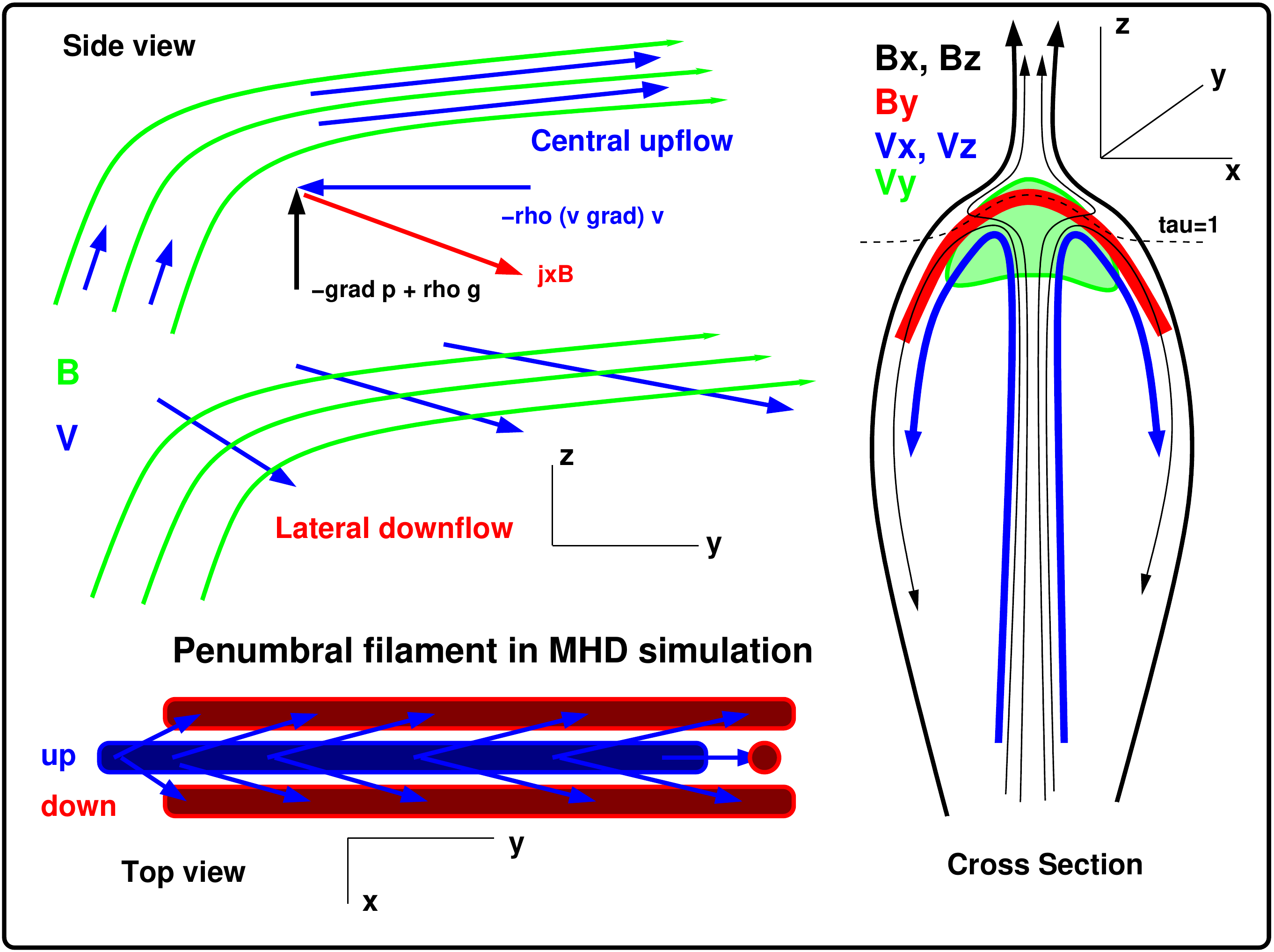}}
  \caption{A sketch summarizing the basic field and flow structure of a 
    penumbral filament as present in the numerical simulation. We present a 
    schematic side, top and cross-section view. $x$ and $z$ denote the 
    horizontal and vertical direction perpendicular to the filament, $y$ the 
    direction along the filament away from the spot center. In the central 
    upflow regions flow and field are well aligned, while the flow submerges 
    mostly horizontal field lines in lateral downflow regions. Overall 
    filaments have a reduced field strength, but they contain a core with a
    non-vanishing vertical field component. Some of the associated flux
    continues upward, some of the flux returns downward within the filament
    cavity. Depending on the position of the $\tau=1$ level the latter might
    become visible as inverse polarity flux. The strong subsurface shear of
    the Evershed flow induces a strong horizontal field component that is
    concentrated along the $\tau=1$ surface. This leads to strongly magnetized
    Evershed flow channels in the visible layer, while the field strength is
    significantly reduced in the subsurface layers.}
  \label{fig:f21}
\end{figure*}

\section{Conclusions}
\label{sect:conclusions}
In deepening our previous studies in \citet{Rempel:etal:2009,
Rempel:etal:Science,Rempel:2011:IAU,Rempel:2011} 
we investigated how boundary conditions 
and grid resolution affect the structure of penumbrae in simulated sunspots
and worked out observational consequences that can help to further constrain
the magneto-convective structure of the penumbra. The main findings are the
following:
\begin{enumerate}
\item[$\bullet$] The magnetic top boundary condition appears to determine the 
radial extent of the penumbra. We do not get an extended
penumbra if we impose a potential field subject to horizontal periodicity.
Thus, simulations in periodic domains cannot predict the extent of penumbrae 
from first principles at this point.
\item[$\bullet$] Magneto convection forming penumbral fine structure and
driving the Evershed flow is robust over the explored resolution range. While
all explored grid resolutions reproduce basic aspects of a penumbra, a 
resolution of at least $48 [24]$~km (horizontal [vertical]) is required 
for a quantitative agreement.
\item[$\bullet$] The overall amount of convective motions in the photosphere
as characterized by the vertical rms velocity shows little dependence on
resolution and extent of the penumbra. It is linked to the intensity through
an approximate relation of the form $I\propto\sqrt{v_{z}^{\rm rms}(\tau=1)}$.
\item[$\bullet$] We find asymmetries between upflow/downflow filling 
factors and velocities that favor the detection of upflows in the inner and
downflows in the outer penumbra. These asymmetries become more pronounced
at higher resolution. Forward modeling likely overestimates 
the visibility of a convective flow pattern in penumbrae at this point.
\item[$\bullet$] Evershed flow channels are strongly magnetized at 
photospheric levels, values around 1.5-2.5~kG are typical. The tendency for 
enhanced horizontal magnetic field within flow channels by a few 100~G 
compared to the background is a robust feature.
\item[$\bullet$] Opposite polarity flux is present in the penumbra at a level 
of about $10\%$ of the total flux of the sunspot, with no resolution
dependence for an otherwise fixed setup. The overall amount of opposite 
polarity flux increases with the extent of a penumbra from $6-12\%$.
\item[$\bullet$] Integrated over the entire penumbra about $40\%$ of the 
downward directed mass flux is associated with magnetic flux having a polarity 
opposite 
to that of the umbra. We see only a moderate increase of this number with 
resolution and extent of the penumbra. Thus, there should be a significant 
fraction of downflows in the penumbra with the same field polarity as the 
umbra, i.e. a substantial amount of mass flux returns by subduction of upward 
pointing magnetic field rather than by flowing along downward pointing
field lines.
\item[$\bullet$] While strong supersonic downflows are present in the outer 
penumbra (we found up to 15~km$\,$s$^{-1}$ flow amplitude), their overall
contribution to the downward directed mass flux is with $2-3\%$ negligible.
\item[$\bullet$] In the outer penumbra we find a substantial population of
bright downflows that lead to a vanishing $I-v_z$ correlation, while the
correlation in the inner penumbra and surrounding plage region are
comparable (reaching moderate values around $50-60\%$). Nevertheless, energy
is transported in all these regions by convection. Thus, low or moderate
$I-v_z$ correlations cannot be taken as an indication for the absence of
convective energy transport, in particular when magnetic field is present
in the photosphere.  
\item[$\bullet$] Most of the vertical mass flux in the penumbra is balanced
within each radial shell. The unbalanced component of the mass flux
(upflows in the inner and downflows in the outer penumbra) is about $15\%$
of the total overturning mass flux integrated over the whole penumbra. 
\end{enumerate}

We conclude this paper by summarizing the magnetic field and flow structure
of penumbral filaments found in numerical simulations to date in Figure 
\ref{fig:f21}. This sketch is mostly based on the conclusions from this 
paper as well as \citet{Rempel:2011} (see Figure 17 therein).

The side view shows the central upflow regions of filaments where the average
field and average flow are well aligned. The pressure driving in the upflows is 
almost orthogonal to the mostly horizontal acceleration of the Evershed flow,
requiring the Lorentz force to facilitate the energy exchange (the net work by
the Lorentz force is essentially zero). In the lateral downflow regions the 
magnetic field continues to point upward leading to a substantial misalignment 
between flow and field and submergence of field lines. The side view describes 
mostly the situation in the inner penumbra, where we do not find a substantial 
amount of inverse polarity magnetic flux. 

The top view shows the central upflow
region with the adjacent lateral downflows. While mass is moving outward
(Evershed flow) it moves to the edge of filaments where it returns beneath the
solar surface mostly through submergence of field lines. Only near the outer end
of the filament a significant fraction of the mass flux returns through flows
along downward pointing field lines. 

The cross section shows a structure with reduced magnetic field strength in 
the subphotospheric layers that is filled with an overturning convection pattern
maintaining the penumbral brightness. While this structure has a lot of
similarity with the idealized field free gap model of 
\citet{Spruit:Scharmer:2006,Scharmer:Spruit:2006}, it is not 
field free. The filament has a core with a substantial vertical field 
component (see also Figure 17 of \citet{Rempel:2011}) that 
is spreading out laterally near the top of the filament.
A fraction of it passes through the curved $\tau=1$ surface (indicated by 
a dashed line) and continues upward. Overturning convection 
can bend some of these field lines back downward, leading to opposite polarity 
flux along the side of filaments. If the $\tau=1$ level is located deep enough,
this can lead to the appearance of opposite polarity flux along the edge of 
filaments in magnetograms. 
The Evershed flow (shown in green) increases rapidly toward the $\tau=1$ level 
in a region with a substantial vertical field component, leading to induction 
of a strong horizontal field component along the axis of the filament.
Overturning motions distribute the magnetic field along the $\tau=1$ level in 
a thin boundary layer shown in red, resulting in strongly magnetized
Evershed flow channels at photospheric levels that might appear like 
horizontal flux tubes to the observer. Note that we simplified the
picture in the sketch with regard to the distribution of the horizontal field 
component. Horizontal field is present throughout most of the filament, it is 
however about a factor of $2$ stronger in the boundary layer indicated in the 
sketch and mostly there dynamically relevant (the acceleration of the Evershed 
flow takes place in this boundary layer, see indicated force balance).
   
\acknowledgements
The author thanks M. Kn{\"o}lker for insightful discussions and very valuable 
comments on the manuscript. This research has greatly benefited from discussions
that were held at the International Space Science Institute (ISSI)
in Bern (Switzerland) in February 2010 as part of the international
working group {\em Extracting Information from spectropolarimetric 
observations: comparison of inversion codes}.
The National Center for Atmospheric Research is sponsored by the National 
Science Foundation. Computing resources were provided by the National 
Institute for Computer Sciences (NICS) under grant TG-AST100005. We thank the 
staff at the supercomputing center for their technical support.

\section{Appendix A}
For initialization of our simulations we use an axisymmetric self-similar 
magnetic field configuration. This approach has been used previously for
magnetohydrostatic sunspot models by \citet{Schlueter:Temesvary:1958,
Deinzer:1965}, and more recently by \citet{Schuessler:Rempel:2005}.
\begin{eqnarray}
  B_z(R,z) &=& B_0\, f(\zeta)\, g(z)\,,
\label{bz} \\
\noalign{\vskip 2mm}
  B_R(R,z) &=& -B_0\, {R\over 2}\,f(\zeta)\,g^{\prime}(z)\,.
\label{br}
\end{eqnarray}
Here $R, z$ correspond to the radial and vertical direction in cylindrical
coordinates, $g(z)$ is the vertical magnetic field profile along the symmetry 
axis and $\zeta = R\sqrt{g(z)}$. The function $f(\zeta)$ describes the
height independent radial profile of the vertical field component and
can be freely chosen. For initializing our simulations we used the following
functions to describe the radial and vertical field profile:
\begin{eqnarray}
  f(\zeta) &=& \exp[-(\zeta/R_0)^4]\, \\
  g(z)&=& \exp[-(z/z_0)^2]\,,
\end{eqnarray}
leading to a total of three parameters that determine the field structure:
$B_0$, $R_0$, and $z_0$. The total flux is given by $\Phi_0=2.7833 B_0 R_0^2$.
Note that the most important parameters for our problem are
$\Phi_0$ and $B_0$, the detailed choices of $f(\zeta)$ and $g(z)$ are
secondary. We have chosen $g(z)$ in this form to ensure $g^{\prime}(z=0)=0$
for consistency with our vertical field bottom boundary condition and 
$f(\zeta)$ to initially concentrate most field into the sunspot. The
typically used Gaussian profile would lead to a smaller sunspot and
a stronger plage region.

We insert the above magnetic field structure into a thermally relaxed HD 
simulation. We do not correct the thermal structure to account for magnetic
forces, instead we allow this adjustment to happen as part of the the 
dynamical evolution. We set however in regions with $B>2$ kG the entropy
to the value found at the bottom of the domain, while keeping the pressure
unchanged and initially quench velocities in magnetized regions. The latter
two have a rather minor effect on the overall evolution.

\section{Appendix B}
For the simulations presented in this publication we used a modified top 
boundary condition, which allows us to arbitrarily to change the inclination 
angle of the magnetic field, while maintaining horizontal periodicity. To 
explain this boundary condition we start from a
potential field extrapolation in a horizontally periodic domain, which can 
be derived from a potential of the form \citep{Cheung:2006:PhD}:
\begin{equation}
\Phi(x,y,z)=\int \tilde{\Phi}(k_x,k_y)e^{-\vert k\vert z} 
e^{i k_x x}e^{i k_y y} dk_x\, dk_y\;,
\end{equation}
with $\vert k\vert=\sqrt{k_x^2+k_y^2}$. The magnetic field components are 
given by $\vec{B}=-\nabla \Phi$, together with the relation 
$\tilde{B}_0(k_x,k_y)=\vert k\vert \tilde{\Phi}(k_x,k_y)$. Here $\tilde{B}_0$
is the Fourier transform of $B_z(x,y,0)$ at the boundary $z=0$.
\begin{eqnarray}
  B_x &=&-\int \tilde{B}_0(k_x,k_y) \frac{i k_x}{\vert k\vert}
  e^{-\vert k\vert z} e^{i k_x x}e^{i k_y y} dk_x\, dk_y\nonumber \\
  B_y &=&-\int \tilde{B}_0(k_x,k_y) \frac{i k_y}{\vert k\vert}
  e^{-\vert k\vert z} e^{i k_x x}e^{i k_y y} dk_x\, dk_y\nonumber \\
  B_z &=&+\int \tilde{B}_0(k_x,k_y)
  e^{-\vert k\vert z} e^{i k_x x}e^{i k_y y} dk_x\, dk_y
\end{eqnarray} 
From this potential field extrapolation we derive a generalized non-potential 
field boundary condition by introducing an additional parameter $\alpha$ 
controlling the field inclination angle:
\begin{eqnarray}
  B_x^{\alpha} &=&-\alpha \int \tilde{B}_0(k_x,k_y) \frac{i k_x}{\vert k\vert}
  e^{-\alpha\vert k\vert z} e^{i k_x x}e^{i k_y y} dk_x\, dk_y\nonumber \\
  B_y^{\alpha} &=&-\alpha\int \tilde{B}_0(k_x,k_y) \frac{i k_y}{\vert k\vert}
  e^{-\alpha\vert k\vert z} e^{i k_x x}e^{i k_y y} dk_x\, dk_y\nonumber \\
  B_z^{\alpha} &=&+\int \tilde{B}_0(k_x,k_y)
  e^{-\alpha\vert k\vert z} e^{i k_x x}e^{i k_y y} dk_x\, dk_y
\end{eqnarray}
While using a value of $\alpha=1$ reproduces the potential field, a value
of $0\leq\alpha<1$ leads to a field more vertical than potential (vertical
field for $\alpha=0$), a value of $\alpha>1$ to a field extrapolation more 
horizontal than potential. 

The electrical current distribution associated with this field extrapolation
is given by:
\begin{eqnarray}
  j_x^{\alpha} &=&+(1-\alpha^2) \int \tilde{B}_0(k_x,k_y) i k_y
  e^{-\alpha\vert k\vert z} e^{i k_x x}e^{i k_y y} dk_x\, dk_y\nonumber \\
  j_y^{\alpha} &=&-(1-\alpha^2) \int \tilde{B}_0(k_x,k_y) i k_x
  e^{-\alpha\vert k\vert z} e^{i k_x x}e^{i k_y y} dk_x\, dk_y\nonumber \\
  j_z^{\alpha} &=& 0
\end{eqnarray}
While the vertical current remains zero in the region $z>0$, closed
horizontal current loops induce magnetic field which changes the effective
field inclination in $z\le 0$. The magnetic field in the region
$z>0$ is not force free, but the magnetic field in the computational domain
$z<0$ remains close to force free due to the low $\beta$ conditions we 
consider.


\begin{thebibliography}{63}
\expandafter\ifx\csname natexlab\endcsname\relax\def\natexlab#1{#1}\fi

\bibitem[{{Bellot Rubio} {et~al.}(2010){Bellot Rubio}, {Schlichenmaier}, \&
  {Langhans}}]{BellotRubio:etal:2010}
{Bellot Rubio}, L.~R., {Schlichenmaier}, R., \& {Langhans}, K. 2010, \apj, 725,
  11

\bibitem[{{Bellot Rubio} {et~al.}(2006){Bellot Rubio}, {Schlichenmaier}, \&
  {Tritschler}}]{Bellot-Rubio:etal:2006}
{Bellot Rubio}, L.~R., {Schlichenmaier}, R., \& {Tritschler}, A. 2006, \aap,
  453, 1117

\bibitem[{{Bharti} {et~al.}(2011){Bharti}, {Sch{\"u}ssler}, \&
  {Rempel}}]{Bharti:etal:2011:conv}
{Bharti}, L., {Sch{\"u}ssler}, M., \& {Rempel}, M. 2011, \apj, 739, 35

\bibitem[{{Bharti} {et~al.}(2010){Bharti}, {Solanki}, \&
  {Hirzberger}}]{Bharti:etal:2010}
{Bharti}, L., {Solanki}, S.~K., \& {Hirzberger}, J. 2010, \apjl, 722, L194

\bibitem[{{Borrero}(2009)}]{Borrero:2009}
{Borrero}, J.~M. 2009, Science in China G: Physics and Astronomy, 52, 1670

\bibitem[{{Borrero} \& {Ichimoto}(2011)}]{Borrero:Ichimoto:2011}
{Borrero}, J.~M., \& {Ichimoto}, K. 2011, Living Reviews in Solar Physics, 8, 4

\bibitem[{{Borrero} {et~al.}(2008){Borrero}, {Lites}, \&
  {Solanki}}]{Borrero:etal:2008}
{Borrero}, J.~M., {Lites}, B.~W., \& {Solanki}, S.~K. 2008, \aap, 481, L13

\bibitem[{{Borrero} \& {Solanki}(2008)}]{Borrero:Solanki:2008}
{Borrero}, J.~M., \& {Solanki}, S.~K. 2008, \apj, 687, 668

\bibitem[{{Borrero} \& {Solanki}(2010)}]{Borrero:Solanki:2010}
---. 2010, \apj, 709, 349

\bibitem[{{Brummell} {et~al.}(2008){Brummell}, {Tobias}, {Thomas}, \&
  {Weiss}}]{Brummell:etal:2008}
{Brummell}, N.~H., {Tobias}, S.~M., {Thomas}, J.~H., \& {Weiss}, N.~O. 2008,
  \apj, 686, 1454

\bibitem[{{Cheung}(2006)}]{Cheung:2006:PhD}
{Cheung}, M. 2006, PhD thesis, University of G{\"o}ttingen, Germany,
  http://webdoc.sub.gwdg.de/diss/2006/cheung

\bibitem[{{Cheung} {et~al.}(2010){Cheung}, {Rempel}, {Title}, \&
  {Sch{\"u}ssler}}]{Cheung:etal:2010}
{Cheung}, M.~C.~M., {Rempel}, M., {Title}, A.~M., \& {Sch{\"u}ssler}, M. 2010,
  \apj, 720, 233

\bibitem[{{Deinzer}(1965)}]{Deinzer:1965}
{Deinzer}, W. 1965, \apj, 141, 548

\bibitem[{{del Toro Iniesta} {et~al.}(2001){del Toro Iniesta}, {Bellot Rubio},
  \& {Collados}}]{DelToro:etal:2001}
{del Toro Iniesta}, J.~C., {Bellot Rubio}, L.~R., \& {Collados}, M. 2001,
  \apjl, 549, L139

\bibitem[{{Deng} {et~al.}(2005){Deng}, {Liu}, {Yang}, {Wang}, \&
  {Denker}}]{Deng:etal:2005}
{Deng}, N., {Liu}, C., {Yang}, G., {Wang}, H., \& {Denker}, C. 2005, \apj, 623,
  1195

\bibitem[{{Franz}(2011)}]{Franz:2011}
{Franz}, M. 2011, ArXiv e-prints

\bibitem[{{Franz} \& {Schlichenmaier}(2009)}]{Franz:Schlichenmaier:2009}
{Franz}, M., \& {Schlichenmaier}, R. 2009, \aap, 508, 1453

\bibitem[{{Heinemann} {et~al.}(2007){Heinemann}, {Nordlund}, {Scharmer}, \&
  {Spruit}}]{Heinemann:etal:2007}
{Heinemann}, T., {Nordlund}, {\AA}., {Scharmer}, G.~B., \& {Spruit}, H.~C.
  2007, \apj, 669, 1390

\bibitem[{{Ichimoto} {et~al.}(2007{\natexlab{a}}){Ichimoto}, {Shine}, {Lites},
  {Kubo}, {Shimizu}, {Suematsu}, {Tsuneta}, {Katsukawa}, {Tarbell}, {Title},
  {Nagata}, {Yokoyama}, \& {Shimojo}}]{Ichimoto:etal:2007}
{Ichimoto}, K., {et~al.} 2007{\natexlab{a}}, \pasj, 59, 593

\bibitem[{{Ichimoto} {et~al.}(2007{\natexlab{b}}){Ichimoto}, {Suematsu},
  {Tsuneta}, {Katsukawa}, {Shimizu}, {Shine}, {Tarbell}, {Title}, {Lites},
  {Kubo}, \& {Nagata}}]{Ichimoto:etal:2007:sc}
---. 2007{\natexlab{b}}, Science, 318, 1597

\bibitem[{{Ichimoto} {et~al.}(2008){Ichimoto}, {Tsuneta}, {Suematsu},
  {Katsukawa}, {Shimizu}, {Lites}, {Kubo}, {Tarbell}, {Shine}, {Title}, \&
  {Nagata}}]{Ichimoto:etal:2008}
---. 2008, \aap, 481, L9

\bibitem[{{Joshi} {et~al.}(2011){Joshi}, {Pietarila}, {Hirzberger}, {Solanki},
  {Aznar Cuadrado}, \& {Merenda}}]{Joshi:etal:2011}
{Joshi}, J., {Pietarila}, A., {Hirzberger}, J., {Solanki}, S.~K., {Aznar
  Cuadrado}, R., \& {Merenda}, L. 2011, \apjl, 734, L18

\bibitem[{{Katsukawa} \& {Jur{\v c}{\'a}k}(2010)}]{Katsukawa:Jurcak:2010}
{Katsukawa}, Y., \& {Jur{\v c}{\'a}k}, J. 2010, \aap, 524, A20

\bibitem[{{Kitiashvili} {et~al.}(2009){Kitiashvili}, {Kosovichev}, {Wray}, \&
  {Mansour}}]{Kitiashvili:etal:2009}
{Kitiashvili}, I.~N., {Kosovichev}, A.~G., {Wray}, A.~A., \& {Mansour}, N.~N.
  2009, \apjl, 700, L178

\bibitem[{{Langhans} {et~al.}(2007){Langhans}, {Scharmer}, {Kiselman}, \&
  {L{\"o}fdahl}}]{Langhans:etal:2007}
{Langhans}, K., {Scharmer}, G.~B., {Kiselman}, D., \& {L{\"o}fdahl}, M.~G.
  2007, \aap, 464, 763

\bibitem[{{Langhans} {et~al.}(2005){Langhans}, {Scharmer}, {Kiselman},
  {L{\"o}fdahl}, \& {Berger}}]{Langhans:etal:2005}
{Langhans}, K., {Scharmer}, G.~B., {Kiselman}, D., {L{\"o}fdahl}, M.~G., \&
  {Berger}, T.~E. 2005, \aap, 436, 1087

\bibitem[{{Liu} {et~al.}(2005){Liu}, {Deng}, {Liu}, {Falconer}, {Goode},
  {Denker}, \& {Wang}}]{Liu:etal:2005}
{Liu}, C., {Deng}, N., {Liu}, Y., {Falconer}, D., {Goode}, P.~R., {Denker}, C.,
  \& {Wang}, H. 2005, \apj, 622, 722

\bibitem[{{M{\'a}rquez} {et~al.}(2006){M{\'a}rquez}, {S{\'a}nchez Almeida}, \&
  {Bonet}}]{Marquez:etal:2006}
{M{\'a}rquez}, I., {S{\'a}nchez Almeida}, J., \& {Bonet}, J.~A. 2006, \apj,
  638, 553

\bibitem[{{Montesinos} \& {Thomas}(1997)}]{Montesinos:Thomas:1997}
{Montesinos}, B., \& {Thomas}, J.~H. 1997, \nat, 390, 485

\bibitem[{{Nordlund} \& {Scharmer}(2010)}]{Nordlund:Scharmer:2010}
{Nordlund}, {\AA}., \& {Scharmer}, G.~B. 2010, in Magnetic Coupling between the
  Interior and Atmosphere of the Sun, ed. {S.~S.~Hasan \& R.~J.~Rutten},
  243--254

\bibitem[{{Rempel}(2011{\natexlab{a}})}]{Rempel:2011:IAU}
{Rempel}, M. 2011{\natexlab{a}}, in IAU Symposium, Vol. 273, IAU Symposium, ed.
  D.~P. {Choudhary} \& K.~G. {Strassmeier} (Cambridge/UK: Cambridge Univ.
  Press), 8--14

\bibitem[{{Rempel}(2011{\natexlab{b}})}]{Rempel:2011}
{Rempel}, M. 2011{\natexlab{b}}, \apj, 729, 5

\bibitem[{{Rempel}(2011{\natexlab{c}})}]{Rempel:2011:moat}
---. 2011{\natexlab{c}}, \apj, 740, 15

\bibitem[{{Rempel} \& {Schlichenmaier}(2011)}]{Rempel:Schlichenmaier:2011}
{Rempel}, M., \& {Schlichenmaier}, R. 2011, Living Reviews in Solar Physics, 8,
  3

\bibitem[{{Rempel} {et~al.}(2009{\natexlab{a}}){Rempel}, {Sch{\"u}ssler},
  {Cameron}, \& {Kn{\"o}lker}}]{Rempel:etal:Science}
{Rempel}, M., {Sch{\"u}ssler}, M., {Cameron}, R.~H., \& {Kn{\"o}lker}, M.
  2009{\natexlab{a}}, Science, 325, 171

\bibitem[{{Rempel} {et~al.}(2009{\natexlab{b}}){Rempel}, {Sch{\"u}ssler}, \&
  {Kn{\"o}lker}}]{Rempel:etal:2009}
{Rempel}, M., {Sch{\"u}ssler}, M., \& {Kn{\"o}lker}, M. 2009{\natexlab{b}},
  \apj, 691, 640

\bibitem[{{Rimmele}(2008)}]{Rimmele:2008}
{Rimmele}, T. 2008, \apj, 672, 684

\bibitem[{{Rimmele} \& {Marino}(2006)}]{Rimmele:Marino:2006}
{Rimmele}, T., \& {Marino}, J. 2006, \apj, 646, 593

\bibitem[{{S{\'a}nchez Almeida}(2005)}]{Sanchez-Almeida:2005:sunspot}
{S{\'a}nchez Almeida}, J. 2005, \apj, 622, 1292

\bibitem[{{S{\'a}nchez Almeida} {et~al.}(2007){S{\'a}nchez Almeida},
  {M{\'a}rquez}, {Bonet}, \& {Dom{\'{\i}}nguez
  Cerde{\~n}a}}]{Sanchez-Almeida:etal:2007}
{S{\'a}nchez Almeida}, J., {M{\'a}rquez}, I., {Bonet}, J.~A., \&
  {Dom{\'{\i}}nguez Cerde{\~n}a}, I. 2007, \apj, 658, 1357

\bibitem[{{Scharmer} {et~al.}(2002){Scharmer}, {Gudiksen}, {Kiselman}, {L{\"
  o}fdahl}, \& {Rouppe van der Voort}}]{Scharmer:etal:2002}
{Scharmer}, G.~B., {Gudiksen}, B.~V., {Kiselman}, D., {L{\" o}fdahl}, M.~G., \&
  {Rouppe van der Voort}, L.~H.~M. 2002, \nat, 420, 151

\bibitem[{{Scharmer} \& {Henriques}(2011)}]{Scharmer:Henriques:2011}
{Scharmer}, G.~B., \& {Henriques}, V.~M.~J. 2011, ArXiv e-prints

\bibitem[{{Scharmer} {et~al.}(2011){Scharmer}, {Henriques}, {Kiselman}, \& {de
  la Cruz Rodr{\'{\i}}guez}}]{Scharmer:etal:2011:sci}
{Scharmer}, G.~B., {Henriques}, V.~M.~J., {Kiselman}, D., \& {de la Cruz
  Rodr{\'{\i}}guez}, J. 2011, Science, 333, 316

\bibitem[{{Scharmer} {et~al.}(2007){Scharmer}, {Langhans}, {Kiselman}, \&
  {L{\"o}fdahl}}]{Scharmer:etal:2007}
{Scharmer}, G.~B., {Langhans}, K., {Kiselman}, D., \& {L{\"o}fdahl}, M.~G.
  2007, in Astronomical Society of the Pacific Conference Series, Vol. 369, New
  Solar Physics with Solar-B Mission, ed. {K.~Shibata, S.~Nagata, \&
  T.~Sakurai}, 71

\bibitem[{{Scharmer} {et~al.}(2008){Scharmer}, {Nordlund}, \&
  {Heinemann}}]{Scharmer:etal:2008}
{Scharmer}, G.~B., {Nordlund}, {\AA}., \& {Heinemann}, T. 2008, \apjl, 677,
  L149

\bibitem[{{Scharmer} \& {Spruit}(2006)}]{Scharmer:Spruit:2006}
{Scharmer}, G.~B., \& {Spruit}, H.~C. 2006, \aap, 460, 605

\bibitem[{{Schl{\" u}ter} \& {Temesv{\' a}ry}(1958)}]{Schlueter:Temesvary:1958}
{Schl{\" u}ter}, A., \& {Temesv{\' a}ry}, S. 1958, in IAU Symp. 6:
  Electromagnetic Phenomena in Cosmical Physics, ed. B.~{Lehnert} (Cambridge
  University Press), 263

\bibitem[{{Schlichenmaier} {et~al.}(2005){Schlichenmaier}, {Bellot Rubio}, \&
  {Tritscher}}]{Schlichenmaier:etal:2005}
{Schlichenmaier}, R., {Bellot Rubio}, L.~R., \& {Tritscher}, A. 2005,
  Astronomische Nachrichten, 326, 301

\bibitem[{{Sch{\"u}ssler} \& {Rempel}(2005)}]{Schuessler:Rempel:2005}
{Sch{\"u}ssler}, M., \& {Rempel}, M. 2005, \aap, 441, 337

\bibitem[{{Sch{\"u}ssler} \& {V{\"o}gler}(2006)}]{Schuessler:Voegler:2006}
{Sch{\"u}ssler}, M., \& {V{\"o}gler}, A. 2006, \apjl, 641, L73

\bibitem[{{Shimizu} {et~al.}(2012){Shimizu}, {Ichimoto}, \&
  {Suematsu}}]{Shimizu:etal:2012}
{Shimizu}, T., {Ichimoto}, K., \& {Suematsu}, Y. 2012, \apjl, 747, L18

\bibitem[{{Solanki}(2003)}]{Solanki:2003}
{Solanki}, S.~K. 2003, {\aap}r, 11, 153

\bibitem[{{Spruit} \& {Scharmer}(2006)}]{Spruit:Scharmer:2006}
{Spruit}, H.~C., \& {Scharmer}, G.~B. 2006, \aap, 447, 343

\bibitem[{{Spruit} {et~al.}(2010){Spruit}, {Scharmer}, \&
  {L{\"o}fdahl}}]{Spruit:etal:2010}
{Spruit}, H.~C., {Scharmer}, G.~B., \& {L{\"o}fdahl}, M.~G. 2010, \aap, 521,
  A72

\bibitem[{{Steiner} {et~al.}(2010){Steiner}, {Franz}, {Bello Gonz{\'a}lez},
  {Nutto}, {Rezaei}, {Mart{\'{\i}}nez Pillet}, {Bonet Navarro}, {del Toro
  Iniesta}, {Domingo}, {Solanki}, {Kn{\"o}lker}, {Schmidt}, {Barthol}, \&
  {Gandorfer}}]{Steiner:etal:2010}
{Steiner}, O., {et~al.} 2010, \apjl, 723, L180

\bibitem[{{Thomas}(2010)}]{Thomas:2010}
{Thomas}, J.~H. 2010, in Magnetic Coupling between the Interior and Atmosphere
  of the Sun, ed. {S.~S.~Hasan \& R.~J.~Rutten}, 229--242

\bibitem[{{Thomas} \& {Montesinos}(1993)}]{Thomas:Montesinos:1993}
{Thomas}, J.~H., \& {Montesinos}, B. 1993, \apj, 407, 398

\bibitem[{{Thomas} \& {Weiss}(2004)}]{Thomas:Weiss:2004}
{Thomas}, J.~H., \& {Weiss}, N.~O. 2004, \araa, 42, 517

\bibitem[{{Thomas} \& {Weiss}(2008)}]{Thomas:Weiss:2008}
---. 2008, {Sunspots and Starspots}, ed. {Thomas, J.~H.~\& Weiss, N.~O.}
  (Cambridge University Press)

\bibitem[{{Tritschler} {et~al.}(2007){Tritschler}, {M{\"u}ller},
  {Schlichenmaier}, \& {Hagenaar}}]{Tritschler:etal:2007}
{Tritschler}, A., {M{\"u}ller}, D.~A.~N., {Schlichenmaier}, R., \& {Hagenaar},
  H.~J. 2007, \apjl, 671, L85

\bibitem[{{V{\" o}gler} {et~al.}(2005){V{\" o}gler}, {Shelyag}, {Sch{\"
  u}ssler}, {Cattaneo}, {Emonet}, \& {Linde}}]{Voegler:etal:2005}
{V{\" o}gler}, A., {Shelyag}, S., {Sch{\" u}ssler}, M., {Cattaneo}, F.,
  {Emonet}, T., \& {Linde}, T. 2005, \aap, 429, 335

\bibitem[{{Westendorp Plaza} {et~al.}(1997){Westendorp Plaza}, {del Toro
  Iniesta}, {Ruiz Cobo}, {Martinez Pillet}, {Lites}, \&
  {Skumanich}}]{Westendorp:etal:1997}
{Westendorp Plaza}, C., {del Toro Iniesta}, J.~C., {Ruiz Cobo}, B., {Martinez
  Pillet}, V., {Lites}, B.~W., \& {Skumanich}, A. 1997, \nat, 389, 47

\bibitem[{{Zakharov} {et~al.}(2008){Zakharov}, {Hirzberger}, {Riethm{\"u}ller},
  {Solanki}, \& {Kobel}}]{Zakharov:etal:2008}
{Zakharov}, V., {Hirzberger}, J., {Riethm{\"u}ller}, T.~L., {Solanki}, S.~K.,
  \& {Kobel}, P. 2008, \aap, 488, L17

\end{thebibliography}
\end{document}